\documentclass[prc,aps,eqsecnum,epsf,amssymb]{revtex4}
\usepackage{graphicx}

\newcommand{\bmq}{{\mbox{\boldmath $q$}}}

\newcommand{\qq}{{|\mbox{\boldmath $q$}|}}

\begin{document}
\preprint {WIS 04/Sept, 21 DPP}
\date{\today}
\title{Inclusive scattering data on light nuclei as a tool for the extraction 
of $G_M^n$ }
\author{A.S. Rinat and M.F. Taragin}
\address{Weizmann Institute of Science, Department of Particle Physics,
Rehovot 76100, Israel}
\author{M. Viviani}
\address{INFN, Sezione Pisa and Phys. Dept., University of Pisa, I-56100,
Italy}

\begin{abstract}

We demonstrate that refinements in the analysis of inclusive scattering data
on light nuclei enable the extraction of, generally accurate, values of 
the neutron magnetic form factor $G_M^n(Q^2)$. In particular, a recent 
parametrization of $ep$ inclusive resonance excitation enables a reliable 
calculation of the inelastic background, and as a consequence a separation 
of quasi-elastic and  inelastic contributions.  
A far larger number of data points than previously considered is now
available for analysis and enables a more reliable extraction of $G_M^n$ 
from cross section and  $R_T$ data on D and He. The achieved 
accuracy appears mainly limited by the present uncertainties in the 
knowledge of proton form factors and by the accuracy of the data.

\end{abstract}

\maketitle

\section{Introduction.}

In a previous report we discussed the feasibility to extract the neutron 
magnetic form factor (FF) $G_M^n(Q^2)$ from inclusive electron scattering 
data on D and $^4$He \cite{rtv} at small and moderate $Q^2$. Using 
a simple model for inclusive resonance excitation, we established, that 
in the immediate vicinity of the quasi-elastic peak (QEP), the QE components 
dominate the inelastic background, which was subsequently neglected. 
That procedure severely restricts the number of data points around  the QEP
available for the extraction of $\alpha_n=G_M^n/\mu_n G_d$
($\mu_n$ and $G_d(Q^2)=[1+Q^2/0.71]^{-2}$ are the static 
magnetic moment of the neutron, and the standard dipole form factor).

For several reasons we propose to extend the above analysis:

1) Availability of an unpublished set of  low-$Q^2$  quasi-elastic (QE)
data on $^4$He, \cite{chen1}, which are contained in a PhD Thesis of 
J-P. Chen \cite{chen2} and which cover a nearly continuous range 
in $Q^2$. Moreover, contrary to the older NE3 data \cite{ne3}, most of the
NE9 data sets reach, or extend into the resonance region.

2) After completion of Ref. \onlinecite{rtv} we have been informed of recent
high-quality inclusive resonance excitation data on a proton, which supersede 
parameterizations of older SLAC \cite{stuart} and of more recent JLab 
data \cite{christy} (see also Ref. \cite{osip}). In addition a model 
has to be devised in order to obtain the required neutron structure functions
(SF) $F_k^n$ ($k=1,2$) \cite{rt1}, which appears in the desired $F_k^N$, the
averaged $p,n$ SF. The latter is related to the nuclear SF $F_k^A$ 
by means of $f^{PN,A}$, the SF of a fictitious nucleus composed of 
point-nucleons, which has to be computed. 

The above  $F_k^N$ may be
separated into nucleon-elastic (NE) and nucleon-inelastic (NI) components 
$F_k^{N,NE}, F_k^{N,NI}$, which correspond to processes where, after 
absorption of a virtual photon, the nucleon is not, or gets excited. 
The above mentioned relation  generates  a similar division  of the 
nuclear SF $F_k^A=F_k^{A,NI}+F_k^{A,NE}$.

In each data set we consider two regions:

A) The high energy-loss region far from the QEP, which is 
dominated by the nuclear NI background: Tails of QE contributions 
barely depend on the precise value of $G_M^n$ and, the overwhelmingly
inelastic computed cross sections can thus be compared against data. 

B) With decreasing energy loss one reaches the inelastic and elastic sides 
of the QEP. There we focus on the difference between data and the NI 
components. To those we apply a criterion which, when fulfilled, identifies 
the above differences as $F_k^{A,NE}$, which is related to the nucleon FFs. 
The QE region will be shown to be the most sensitive one for variations in 
$\alpha_n$, and is hence the area of prime interest.

The above program is hampered by complicating circumstances. Foremost is 
the  discrepancy \cite{arr2,arr3} between the $G_E^p/G_M^p$ ratio, extracted
from Rosenbluth-separated elastic $ep$ cross sections and from  
${\vec e}(p,{\vec p})e$ polarization transfer data \cite{mjon,mjon1}. At 
this time it is not obvious which $p$ FFs should be used. It seems that 
two-photon exchange contributions \cite{blund, wally} confirm the $E/M$ 
ratio for the proton, provided by the polarization transfer data. 
Unfortunately this does not directly reflect on the FFs themselves,
because the $ep$ data have as yet not been corrected for those contributions, 
prior to an extraction. Before resolution of the above issue, a choice for 
the required input will have to be made. Arrington recommends the use of 
the experimentally simplest, non-separated $ep$ cross section data.  

The present note is organized as follows. We start with 
total inclusive cross sections, but shall also re-analyze the components 
for the absorption of virtual transverse photons $\propto R_T$. After 
discussing the required input, we analyze all good-quality  data on 
D, $^4$He. 

We first concentrate on the region below and around the QEP, where NE 
components generally dominate. This enables one to reliably extract 
$\alpha_n$ from any given data set. We no more neglect NI, as had 
previously been done \cite{rtv}, and instead subtract those from the data  
in order to isolate the NE components. We also show that the computed 
cross sections  agree very well with the data in and beyond the resonance 
region, which are completely dominated by NI components.  Between the 
inelastic slope of the QE region and the elastic wing of the resonance 
region, the NI components falls short of their predictions and we 
discuss a possible origin.

In the discussion we set limits to the accuracy of the results, caused 
by occasionally data of insufficient quality, the ambiguity of the 
proton FFs, and uncertainty in the NI components. Without substantial
improvements in each of the above items, we do not envisage the possibility
to substantially improve the results of the model.

\section{Quasi-Elastic inclusive scattering.}

\label{sec:gnr}

Consider the cross section per nucleon for inclusive scattering over an
angle $\theta$ of unpolarized electrons with energy $E$
\begin{eqnarray} 
\frac{d^2\sigma^A(E;\theta,\nu)}{d\Omega\,d\nu}
=\sigma_M(E;\theta,\nu)\bigg\lbrack\frac {2xM}{Q^2}
  F_2^A(x,Q^2)+ \frac{2}{M}F_1^A(x,Q^2){\rm tg}^2(\theta/2) \bigg\rbrack,
\label{a1}  
\end{eqnarray}
with $\sigma_M$, the Mott cross section. Above, $F_{1,2}^A(x,Q^2)$ are the 
nuclear SF, depending on the squared 4-momentum transfer 
$q^2=-Q^2=-(\qq^2-\nu^2)$ and the Bjorken variable $x=Q^2/2M\nu$, 
with $M$ the nucleon mass. Its range is $0\le x\le A$. 

In an alternative representation 
\begin{mathletters}
\begin{eqnarray}
\label{a2}
\frac{d^2\sigma^A}{d\Omega\,d\nu}
&=&\sigma_M[W^A_2+2W^A_1{\rm tg}^2(\theta/2)]
\label{a2a}\\
&=&\sigma_M\frac {Q^2}{|\bmq|^2}[W^A_L+\epsilon^{-1}W^A_T]\equiv
\sigma_M\frac{Q^2}{|\bmq|^2}
\bigg [\frac {Q^2}{|\bmq|^2} R_L^A +\frac{1}{2}\epsilon^{-1}R^A_T \bigg],
\label{a2b}
\end{eqnarray}
\end{mathletters}                        
where  $\epsilon^{-1}=1+2(|\bmq|^2/Q^2){\rm tan}^2(\theta/2)$. The SF 
$W^A_L, W^A_T$ (\ref{a2b}) relate to the absorption of longitudinal and 
transverse photons.
        
In  the sequel we shall use a relation between 
SF $F_k^{N,A}$ for nucleons ($N=p,n$) and a nucleus which, for isospin 
$I=0$ targets of our interest, reads \cite{gr} 
\begin{eqnarray}                                                 
  F^A_k(x,Q^2)=\int_x^A\frac {dz}{z^{2-k}} f^{PN,A}(z,Q^2)\sum_l 
   C_{kl}(z,Q^2) \bigg [F_l^p \bigg (\frac {x}{z},Q^2\bigg )
  +F_l^n\bigg (\frac {x}{z},Q^2\bigg ) \bigg ]\bigg /2                
\label{a3}                                                    
\end{eqnarray}                                    
The link between $F^N=(F^p+F^n)/2$ and $F^A$ is provided by $f^{PN,A}$, the 
SF of a fictitious target $A$, which is composed of point-nucleons. Eq. 
(\ref{a3}) holds in the Bjorken limit $Q^2 \to \infty$, as well as in the 
Plane Wave Impulse Approximation (PWIA). 

For quite some time we have considered Eq. (\ref{a3}) for finite $Q^2$
in an alternative, non-perturbative approach \cite{gr}, 
based on a covariant generalization \cite{gr1} of the non-relativistic 
theory of Gersch-Rodriguez-Smith (GRS) \cite{grs}.  We considered
the above as a conjecture, and its apparent validity for  
$Q^2 \gtrsim Q_0^2\approx$ 2.5 GeV$^2$ as an empirical fact 
\cite{rtval}. 

Only recently did we come across work by West and Jaffe who more than 20
years ago proved Eq. (\ref{a3}) in the PWIA, using either a parton model or 
pQCD \cite{west,jaffe}. It is actually possible to generalize their proof by 
adding Final State Interactions (FSI) to the PWIA, reaching the Distorted 
Wave Impule Approximation. The same holds for the inclusion of FSI in the 
GRS version \cite{rtlink} and the proof formally recovers Eq. (\ref{a3}). 
The intriguing difference lies in the interpretation: in the effective 
hadronic description one uses typical nuclear concepts, as are nuclear 
density matrices, effective $NN$ scattering amplitudes, etc. Those are 
of course foreign concepts in QCD. Similar remarks hold for recently 
discussed effective nuclear parton distribution functions \cite{rt3}. 
 
We return to Eq. (\ref{a3}), which includes the effect of 
mixing of the nucleon SF embodied  in 
the coefficients $C_{kl}$ \cite{atw,sss}. In both the  PWIA and  the GRS 
approach, $C_{11}=1,\,C_{12}=0$, while $C_{21}$ is negligibly small. For a 
discussion of an approximate fashion to compute $C_{22}$ in the GRS, we 
refer to Appendix A of  Ref. \onlinecite{rtv}. Since the approximately
calculated deviation of $C_{22}$ from 1 does not decrease fast enough with 
$Q^2$, we use $C_{22}(Q^2)\to 1$  for $Q^2\gtrsim 3.5\,$GeV$^2$.

Finally we remark that Eq. (\ref{a3}) relates to nucleons as the dominant 
source of partons. This is the case for $x\gtrsim 0.20$ \cite{lle} and thus 
certainly for the range on which we focus $0.4 \lesssim x \lesssim 1.2$,
which comprises the QEP. 

Next we recall the NE and NI parts $F_k^{N,NE}, F_k^{N,NI}$ of nucleon SF, 
which correspond to the elastic absorption of a virtual photon on 
a $N\,$ $\gamma^*+N\to N$  and to inelastic absorption $\gamma^*+ N\to$ 
(hadrons, partons). Elastic components for a nucleon vanish except for $x=1$ 
and contain the standard combinations of the electro-magnetic FF 
$G_{E,M}^N(Q^2)$. Denoting the average of their squares by 
$[{\tilde G}^N]^2=[(G^p)^2+(G^n)^2]/2$, one has ($\eta=Q^2/(4M^2)$) 
\begin{mathletters}
\label{a4}
\begin{eqnarray}
F_1^{N,NE}(x,Q^2)&=&\frac{1}{2}\delta(1-x)[{\tilde G}_M^N(Q^2)]^2
\label{a4a}\\
F_2^{N,NE}(x,Q^2)&=&\delta(1-x) 
\frac {[{\tilde G}_E^N(Q^2)]^2+\eta [{\tilde G}_M^N(Q^2)]^2]}{1+\eta}
\label{a4b}
\end{eqnarray}
\end{mathletters}
It is a trivial matter to express the corresponding NE, NI parts of $nuclear$ 
SF $F_k^A$, using the link (\ref{a3}) in its region of validity 
$x\gtrsim 0.2 $. Thus for the nuclear NE (QE) parts
\begin{mathletters}
\label{a5}
\begin{eqnarray}
  F_1^{A,NE}(x,Q^2)&=&\frac {f^{PN,A}(x,Q^2)}{2}[{\tilde G}_M^N(Q^2)]^2] \,
  \label{a5a}\\
\noalign{\medskip}
  F_2^{A,NE}(x,Q^2)&=&xf^{PN,A}(x,Q^2) C^A_{22}(x,Q^2) 
\frac{[{\tilde G}_E^N(Q^2)]^2+ \eta [{\tilde G}_M^N(Q^2)]^2}{1+\eta}
  \label{a5b}
\end{eqnarray}
\end{mathletters}
Since  $f^{PN,A}$ for light nuclei is sharply peaked around $x\approx 1$, 
the same holds for $F_k^{A,NE}$. For the $L,T$ components corresponding 
to Eqs. (2.5), (2.6) one has
\begin{mathletters}              
\label{a6}
\begin{eqnarray}                 
R_T^{N,NE}(x,Q^2)&=&\delta(1-x)\frac {[{\tilde G}_M^N]^2}{M}
\label{a6a}\\
R_L^{N,NE}(x,Q^2)&=&\delta(1-x)\frac{(1+\eta)[{\tilde G}_E^N]^2}{2M\eta}
\label{a6b}
\end{eqnarray}
\end{mathletters} 
Clearly the NE $L,T$ components for the $N$ separate magnetic and electric 
FF. However, the generalization of the above to composite targets depends 
on the model which relates NE parts of the nucleon and target SF. 
In the GRS approach one finds from  Eqs. (2.4), (2.9) and (2.10)
\begin{mathletters}
\label{a7}
\begin{eqnarray}
R_T^{A,NE}&=&\frac{f^{PN,A}}{M}[{\tilde G}_M^N]^2
\label{a7a}\\
R_L^{A,NE}&=&
\bigg (1+\frac{\eta}{x^2}\bigg )\frac{f^{PN,A}}{2M(1+\eta)}
\bigg \lbrack
\bigg \lbrace C_{22} \bigg (\frac{x^2}{\eta}+1\bigg )\bigg \rbrace
[{\tilde G}_E^N]^2
\nonumber\\
&&~~~~~~~+\bigg \lbrace(C_{22}-1)(x^2+\eta)+x^2-1\bigg \rbrace
[{\tilde G}_M^N]^2 \bigg \rbrack
\label{a7b}
\end{eqnarray}
\end{mathletters}
The NE part of the transverse nuclear SF still contains only magnetic FFs.
However, for the model defined by Eq. (\ref{a3}) the longitudinal nuclear
partner $R_L^{A,NE}$, Eq. (\ref{a7b}) is generally a combination of $E,M$ 
FFs, except at the position $x=1$ of the unshifted QEP, and then only for
unmixed $F_k^A$ in Eq. (2.4), i.e. with $C_{22}=1$. The above contrasts 
with the PWIA, where no such mixture occurs in $R_L^{A,NE}$.
Only in a very limited number of inclusive scattering experiments have
Rosenbluth $L,T$ separations been performed. As a consequence most of
our efforts are concentrated on the more involved total inclusive data.

When focusing on FFs, one has to isolate in the data the NE parts (2.7), 
(2.8), which contain those FFs. Such a procedure obviously requires accurate 
knowledge of the nuclear NI background in the QE region $x\approx 1$, and 
which on the adjacent inelastic side of the QEP $x\lesssim 
x_0(Q^2) \lesssim 1$, is dominated by inclusive resonance excitation. 

Just as the NE components in $F_k^{N,NE}$ produce in nuclei the 
corresponding $F_k^{A,NE}$, which is centered around the QE peak, also 
$N$-resonances in $F_k^{N,NI}$ may be reflected in nuclei SF $F_k^{A,NI}$. 
In particular for $A\le 4$, for which $f^{PN,A}$ is sharply peaked in $x$, 
both $F_k^{N,NI}$ and $F_k^{A,NI}$ have maxima at about the same $x$. We 
recall, that the observed structures in $F_k^{A,NI}$ for $Q^2 \le 3$ GeV$^2$ 
and $\nu>\nu_{\rm QEP}$, are $not$  genuine target resonances but nucleon 
resonances, modified by the nuclear medium \cite{mand}. We refer to those
structures as $'$pseudo-resonances$'$ \cite{hoenig}. 

We conclude this section returning to the validity of Eq. (\ref{a3}).
The simple forms (\ref{a5a}) and (\ref{a5b}) for the nuclear NE parts hold 
quite generally, even below the estimated $Q_0^2\approx 2.5\,$GeV$^2$. The 
same may be assumed for the NI part,  
due to the inclusive excitation of very narrow resonances. With some
hesitation we shall therefore make applications even for $Q^2=0.5\div 1$
GeV$^2$, but one should be prepared to encounter less good fits for those, 
than for larger $Q^2$.

\section {Input.}

We review major theoretical input elements:
  
1) Density matrices for the target nuclei, diagonal in all coordinates
except one. Those relate to ground state wave functions and have for the
lightest nuclei been calculated with great precision \cite{rtd,viv}. 
For heavier targets one has to invoke approximations, for instance by 
interpolation between diagonal density matrices and special limiting 
situations \cite{grs,rt}. Wishing to avoid theoretical uncertainties,
we do not incorporate in our analysis data for targets with $A\ge 12$.

2) $NN$ dynamics for Final State Interactions (FSI) \cite{rt}, which enters 
the calculation of $f^{PN,A}$.

3) $F_k^{p;NI}$: We considered various representations for 
$F_k^p$, all having explicit resonance and background components, namely
Stuart $et\,al$ \cite{stuart} and more recent ones $'$christy1$'$ 
and $'$christy2$'$ \cite{christy}, based on Rosenbluth-separated cross 
sections (cf. also \cite{osip}). The second version is claimed 
to be of somewhat better quality. However, the argument $x/z$ of the latter 
in Eq. (\ref{a3})  varies for fixed $Q^2$ and occasionally crosses stated 
regions of validity. The ensuing inconsistencies appear more severe for
christy2 than for christy1 and we thus  prefer  the latter. Since the 
parameterizations are of relatively poorest quality for small $Q^2$, one 
should expected correspondingly inferior results. 

For large $Q^2$  one avoids uncertainties and even inconsistencies 
by choosing a fixed $Q^2=3.5$ GeV$^2$, beyond which we switch to a 
parametrization of $F_2^p$, averaged over resonances \cite{arn}. For 
$F_1^p$ at high $Q^2$ we employ those of Bodek-Ritchie \cite{bod}. The 
chosen procedure is in line with 
quark-hadron duality, which predicts similar outcome for $F_k^N$, both 
globally, when averaged over the entire resonance region \cite{bg}, and 
locally for isolated resonances \cite{nicu,wally1}. 

4) $F_k^{n;NI}$: We use a procedure presented in Ref. \onlinecite{rt1}, 
based on  the ratio 
\begin{eqnarray}
  {\cal C}(x,Q^2)&=&F_2^n(x,Q^2)/F_2^p(x,Q^2)
  \nonumber\\
  &\equiv&\sum_{k=0}^{2} d_k(Q^2)(1-x)^k,
  \label{a9}
\end{eqnarray}
with the coefficients $d_k(Q^2)$ to be determined by information on 
${\cal C}$ for 3 selected points:

i)  ${\cal C}(0,Q^2)=1$, required to obtain a finite Gottfried sum \cite{rt1}. 

ii)  Use of the primitive choice $F^n=2F^D-F^p$, which is accurate for 
     $x\le 0.3$ and is exploited for the value $x=0.2$.  

iii) Information from the elastic end-point $x=1\,\,$, where ${\cal C}$ is 
determined by  static FF
\begin{eqnarray}
  {\cal C}(1,Q^2)=\frac {\bigg [G_E^n(Q^2)\bigg ]^2+
  \eta \bigg [G_M^n(Q^2)\bigg ]^2}
   {\bigg [G_E^p(Q^2)\bigg ]^2+\eta \bigg [G_M^p(Q^2)\bigg ]^2}\ .
\label{a10}
\end{eqnarray}
We  assume the same ${\cal C}$ for an estimate of $F_1^n$ from $F_1^p$: 
Eq. (\ref{a9}) then provides $F_{k=1,2}^N(x,Q^2)$. The latter and the 
computed $f^{PN,A}$ are from  Eq. (\ref{a3}) seen to be input for the 
calculation of nuclear SF $F_k^A$.

Note that Eq. (\ref{a10}) for ${\cal C}(1,Q^2)$ requires also $G_M^n$, the 
very FFs we wish to extract. Fortunately, for not too large $Q^2$ the ratio 
in Eq. (\ref{a10}) depends only marginally on $G_M^n$. Solely in order to 
determine ${\cal C}(1,Q^2)$, we use in Eq. (\ref{a10}) the parametrization 
of $G_M^n$ given in Ref. \onlinecite{bba}: we do not demand self-consistency. 

5) Electromagnetic FFs: In Ref. \onlinecite{rtv} we had adopted the ratio 
\begin{eqnarray}
\gamma &=& \frac{G_E^p}{(G_M^p\mu_p)}
           \nonumber\\
       &=&1 ~~~~~~~~~~~~~~~~~~~~~~~~~~  {\rm for\ } 
              Q^2\lesssim 0.3\, {\rm GeV}^2\ ,\qquad\qquad\qquad
            {\rm choice\ I}\ ,
           \nonumber\\
       &\approx&[1-0.14(Q^2-0.3)] ~~ {\rm for\ }
              0.3\lesssim Q^2 \lesssim 5.5\, {\rm GeV}^2
\label{a99}
\end{eqnarray}
from  polarization transfer in 
${\vec e}(p,{\vec p})e$ \cite{mjon,mjon1}, $\alpha_p=G_E^p/(\mu_pG_d)$.

For all but the smallest $Q^2$, the above mentioned $E/M$ ratio for 
the $p$ FFs's disagrees substantially from results from  
Rosenbluth-separated elastic $ep$ data \cite{arrros,xxx}. 
Two-photon exchange contributions have been computed \cite{blund}), but 
those have as yet not been extracted from data in order to re-analyze 
the extraction of FFs.

It is therefore impossible at this moment to make an impartial choice 
for $p$ FFs, and we shall report results for two sets of extracted 
$\alpha_n$:  $'$I$'$, based on the ratio (\ref{a99}) and 
$'$II$'$, using cross section data. We adopt the recommended 
parameterizations of $G_{E,M}^p$ \cite{bba}
\begin{eqnarray}
G_{E,M}^p(Q^2)=G_{E,M}^p(0){\bigg /}[1+ \sum_{m=1} a^p_m (Q^2)^m]\ ,
\qquad\qquad\qquad{\rm choice\ II}\ .\label{a11}
\end{eqnarray}
A previously used parametrization for $\alpha_p$  \cite{brash} is close to 
one of the form (\ref{a11}) (see Table 2 in Ref. \onlinecite{bba}).

There is still lacking reliable data for $G_E^n$ beyond relatively low
$Q^2$ \cite{berm}. Analyses continue to prefer a Galster-like form
\begin{eqnarray}
G_E^n(Q^2)=-\mu_nG_d(Q^2)\frac {A\eta}{1+B\eta},
\label{a12}
\end{eqnarray}
with $A=0.942, B=4.61$ \cite{galster,rocco,krutov}. Two recent, more 
precise measurements yield $A=0.888, B=3.21$ \cite{madey}. Differences 
do no amount too much for the small measured $Q^2$: for larger $Q^2$, 
where $G_E^n$ grows in relative importance, there is no way 
to prefer one particular assumed extrapolation. 

\section{General observations and extraction procedure.}

We start with the NE parts of inclusive cross sections and remark on  
two samples for quite different $Q^2$, namely $^4$He 
($E=3.6\,\,{\rm GeV}, \theta=15^{\circ},\,Q^2\,\approx 0.973\,\, 
{\rm GeV}^2)$ and D ($E=4.045\,\,{\rm GeV},\theta=30^{\circ},\,
Q^2\,\approx 2.77\,\, {\rm GeV}^2)$. Figs. 1,2  show on a 
linear scale the cross section data (solid circles), 
the calculated NI background (dotted curves) and $\sigma^{A,NE}$ computed 
for some $\alpha_n$, and two additional values, differing from the 
central one by $\approx 5\%$. Those curves can be compared with 
the difference between the cross section data and the computed NI component
(open circles). The dot-dashed curves are empirical NI components, which
will be discussed below. The following observations from the above 
comparison are not limited to the chosen examples, but hold quite generally:

a)  Whether a QEP stands out in the data or not, in regions away from $x 
\approx 1$,$\sigma^{A,NE}$ is barely modified when $\alpha_n$ is changed by 
as much as 5$\%$ around some average value. Only in the immediate neighborhood 
of the QEP is $\sigma^{A,NE}$ sensitive to small variations in $\alpha_n$.

b) In view of a), a precision extraction of $\sigma^{A,NE}$ in the region 
$x \approx 1$, and thus indirectly of $\alpha_n$, requires a well-determined, 
locally small, inelastic background. The accuracy of such an extraction is  
limited by the precision in the input nucleon SF, the quality of data and of 
the calculated $f^{PN,A}$.

We summarize expectations for inclusive cross sections \cite{rtv}:

i) On the low-$\nu$, elastic side side of the QE region, cross sections for 
small $Q^2$ are predominantly NE and the quality of the extracted $\alpha_n$ 
depends on the precision with which one can calculate the point-nucleon 
nuclear SF  $f^{PN,A}$ in the small wings, away from the peak.

ii) Approaching the QE region, NE components still dominate, provided $Q^2$
is not too large. If $f^{PN,A}$ is sharply peaked, as is the case for the 
lightest nuclei, the same will be observed in
$\sigma^{A,exp}\approx\sigma^{A,NE}$. No matter what the kinematic conditions
are, the QE region close to $x\approx 1$ is most sensitive to $\alpha_n$
and  is therefore the primary source for the looked-for information.

iii) On the inelastic, large-$\nu$ side of the QE region, NE components 
decrease relatively to the increasing inelastic background. 

iv) For further increasing $\nu$, inclusive excitation of $N$ resonances
produces pseudo-resonances in the target. For $f^{PN,A}$ with a sharp maximum,  
pseudo-resonances in nuclei and genuine resonances  
in nucleons, peak at about the same $x$.  For increasing $Q^2$, 
pseudo-resonances  gradually coalesce in a smooth background, due to 
overlapping tails of higher pseudo-resonances.

The extraction procedure follows from the above. First we define reduced 
cross sections on a $p$ and a composite target $A$              
\begin{eqnarray}
  K^{p,A}(E:\theta,\nu)=\frac{d^2\sigma^{ep,eA}}{d\Omega d\nu}\bigg
  /\sigma_M(E;\theta,\nu)=\bigg [\frac{F_2^{p,A}(x,Q^2)}{\nu}
  +\frac {2F_1^{p,A}(x,Q^2)}{M}{\rm tg}^2(\theta/2)\bigg ]\ .
\label{a8}
\end{eqnarray}
Eqs. (2.7) and (2.8) show that $K^{A,NE}(x,Q^2)$ and  $f^{PN,A}$ have 
a similar rapid variation with $x$. 

Next we subtract the computed NI background from the data over the entire 
measured $x(\nu)$-range. We then search for a  continuous $x$ range for 
which the difference of reduced cross sections $\Delta K^A\equiv K^{A,exp}
-K^{A,NI}$ and $f^{PN,A}$ shows maximal similarity. In that range we identify 
\begin{eqnarray}
K^{A,exp}-K^{A,NI}\Longleftrightarrow K^{A,NE}
\label{a43}
\end{eqnarray} 
and determine for each data-point in the selected $x$-range
\begin{mathletters} 
\begin{eqnarray}
\label{a13}         
\alpha_n(Q^2)|\mu_n|&=&\bigg[\frac{2MK^{A,NE}/w^Av^A-B^2/\eta}
{1+{\rm tg}^2(\theta/2)/v^A}\bigg |_{x,Q^2}-
[\alpha_p(Q^2)\mu_p]^2\bigg ]^{1/2}
\label{a13a}\\
&=&\bigg [2MR_T^{A,NE}/w^A\bigg |_{x,Q^2}-[\alpha_p(Q^2)\mu_p]^2\bigg ]^{1/2}
\label{a13b}
\end{eqnarray}
\end{mathletters}
Above $w^A(x,Q^2)=f^{PN,A}(x,Q^2)G_d^2(Q^2);\,v^A(x,Q^2)=
x^2C^A_{22}(x,Q^2)/[2(1+\eta)]$ and $B^2(Q^2)=2[G_E^N(Q^2)]^2/G_d^2(Q^2)$. 

We remark that Eq. (\ref{a3}) also predicts that $K^{A,NE}$ and $f^{PN,A}$ 
have practically the same $A$-dependence. Eqs. (\ref{a13a}), (\ref{a13b}) 
show that residual dependence of $K^{A,NE}$ on both $A$ and $x$ is far 
weaker than the same in $f^{PN,A}$. In addition one notices in Eq. 
(\ref{a13a}) a dependence on $\theta$. For very forward scattering angles 
the above extraction method may occasionally become unstable. An example
will be mentioned below.

Strict fulfillment of the above requirement implies that $\alpha_n$ 
is independent of the data points, which have been selected for the 
extraction. In practice one deals with data for fixed $E,\theta$ and varying 
$x$, hence with $\,\,Q^2$ varying over the measured $\nu$-range. In general 
that variation is mild, but not insignificant for $Q^2\lesssim 0.7-0.8\,$
GeV$^2$, and the extracted $\alpha_n$ will vary there with $Q^2$. But even 
for data sets with fixed  $Q^2$, experimental inaccuracies and the 
imperfections in the theoretical model, cause extracted $\alpha_n$ to depend 
on the selected data points $x_j$. Ultimately one has for each data set to 
determine an average $\langle\alpha_n\rangle$.  

It is virtually impossible to incorporate in the analysis all experimental 
errors and  uncertainties in both input and theory, and we therefore limit 
ourselves to the following:

a) Published tabulated cross sections give statistical and occasionally
systematic errors, but frequently only the former are shown in figures.
Only those are incorporated in our analysis. Most abundant and accurate 
are D data, and occasionally one can assign practically constant 
relative errors  for selected $x$-intervals. 

b) As mentioned in the Introduction the issue of the proton form factors 
is not yet settled. Two-photon exchange contributions to $ep$ inclusive 
scattering apparently influence on the few $\%$ level \cite{blund,wally}. 
Although the $E/M$ ratio for the proton as measured in the polarization 
transfer measurement \cite{mjon,mjon1} is believed to be correct, $p$ FFs,
cannot be determined, without first to apply the above corrections to cross
section data. This has not yet been done. The above reflects on both the 
proton data and the parametrization of Ref. \cite{brash} for $\alpha_n$, 
which we used in Eqs. (\ref{a13a}), (\ref{a13b}). 

At this point we remark that
two-photon exchange contributions cannot be accommodated in a single
generalized convolution (2.4). The observation, that data for $K^{exp}(x)
-K^{NI}(x)$ and $f^{PN,A}(x)$ have very similar variation with $x$, does not 
allow more than a few $\%$ two-photon exchange contributions. 

We therefore stuck to the procedure followed in Ref. \cite{rtv}, using
two sets of proton FFs: one using the Jones results and a second 
one suggested in Ref. \cite{bba}, both not yet corrected for two-photon
exchange contributions. We expect the two methods to provide extremes
between which the correct results will ultimately fall. For both sets 
we applied the error analysis a). 
 
c) The uncertainty in the electric form factor of the neutron appears 
to be of no consequence. Assuming that the different parametrizations may 
be extrapolated to the largest $Q^2$ needed, the ratio $[G_E^n/G_E^p]^2$ 
in $B^2$, Eq. (\ref{a13a}) may vary by as much as $30\%$. However, for
$Q^2 \gtrsim 0.5$ GeV$^2$ the above $B^2$ term is far smaller than the 
first term in the numerator in Eq. (\ref{a13a}), and its inclusion 
affects $\alpha_n$ by less than $1\%$!

We conclude this Section by the following remark. Our extraction 
method for $\langle \alpha_n \rangle$ rests on a test, checking whether
the difference $K^{A,NE}=K^{A,exp}-K^{A,NI}$ and the SF $f^{PN,A}$
have similar $x$-dependence over a continuous set of data points. Having 
determined from those values $\alpha_n(x_i)$ and their average 
$\langle \alpha_n\rangle$, we calculate the corresponding NE component 
NE($\langle \alpha_n\rangle)\equiv\sigma^{A,NE}(\langle \alpha_n \rangle)$.
Likewise NI(comp) stands for the computed $\sigma^{A,NI}$.

The above defined NE($\langle \alpha_n\rangle)$ are constructed to
fit in the mean  NE(extr)=data-NI(comp) in the selected range of data points 
$x_i$ around the QEP. Those actually continues to approximately reproduce 
NE(extr) over a range beyond the chosen points of extraction.
 
Barring the effect of a mildly varying $Q^2$ over the the points of a 
data set, perfect data and an exact theory for NI ought to produce a 
NE($\langle \alpha_n\rangle$), fitting NE(extr) over the $entire\,$ $x$ 
or $\nu$ interval. In the following Section we shall find that deviations 
set in  beyond some $\nu$, where NI about overtakes NE. Those deviations 
reach a maximum around the position of the first pseudo-resonance and 
then rapidly decreases to 0. The culprit may well be NI(comp), in which 
case we define an empirical NI by 
\begin{eqnarray}
  {\rm NI(emp)} \approx {\rm data - NE}(\langle \alpha_n \rangle)
\label{a77}
\end{eqnarray}
By construction  NE$(\langle \alpha_n \rangle)$ closely fits data-NI(emp). 
We shall return to a possible source of the apparent insufficiency of 
NI(comp).

\section{Data and results.}

Below we report on $\alpha_n (G_M^n)$,  extracted from abundant cross 
sections for total inclusive scattering of unpolarized electrons on D 
and $^4$He. Additional information comes from, partly 
re-analyzed sparse data on their transverse components for both targets.

We start with particulars on data and results collected in Table I. Columns
refer to target, beam energy, scattering angle, range of $x,Q^2$ and the
value ${\bar Q}^2\equiv Q^2(\nu_{QEP})$ at the QEP. In the last column we 
first mention the number of  selected data points $x_j$ for each data set. 
Those are followed by the weighted averages of the extracted $\alpha_n$ 
with their error of the mean, which includes statistical errors on the cross 
sections for both parameterizations I and II discussed in Section III. Since 
systematic errors have been disregarded, the stated error bars
should be considered as lower limits. 

Only a sample of analyzed data sets are presented in Figs. 1-12. The two 
options I, II for $p$ FFS produce the same elastic $ep$ cross sections 
with different E/M ratios. For that reason there is no need to specify 
the option in Figs.: it enters only in the ultimately extracted $\alpha_n$. 

HE1): $\,\,E=2.02$ GeV, $\theta=20^{\circ}$; 
$E=3.595$ GeV, $\theta=16,20^{\circ}$; 
${\bar Q}^2= 0.434, 0.873, 1.270$ GeV$^2$ \cite{ne3}.

The above NE3 data  have previously been analyzed \cite{rtv} and the
very good fits on the elastic side of the QEP have been attributed to 
accurate knowledge of the required density matrices and the inclusion of 
$C_{22}$. The present refined calculations produce the observed rise of 
the first two cross sections towards the pseudo-resonances. Those are 
locally $\approx 20\%$ short of the first two data sets and $\approx 10\%$ 
for the third one. Fig. 3 illustrates the latter one.  Since the involved 
$Q^2$ are relatively small, FFs for both options I, II are essentially the 
same. 
                
HE2):  $\,\,E=2.7,\, 3.3,\, 3.6,\, 3.9,\, 4.3\,$GeV, $\theta=15^{\circ}$,
${\bar Q}^2= 0.453, 0.662, 0.781, 0.907, 1.090\,$GeV$^2$; 
$E=0.9, 1.1$ GeV, $\theta=85^{\circ}$,
${\bar Q}^2= 0.78, 1.09\,$GeV$^2$.  \cite{chen1,chen2}. 

The above NE9 cross section data are unpublished parts of the PhD. Thesis 
of J-P Chen \cite{chen1}. Those are in principle a welcome source of 
information on $\alpha_n$ over a dense $Q^2$-range, which partly overlap 
the $Q^2$ range of the the NE3 data, but extend beyond the QEP and the 
adjacent minimum, and occasionally into the pseudo-resonance region. 

A comparison with data, illustrated by Fig. 4 for $E=4.3$ GeV,\,$\theta=
15^{\circ}$ shows a pattern, similar to that for the NE3 data. There is a 
deficiency of $\approx 30 \%$ at the peak of the first pseudo-resonance for 
the set with lowest $Q^2$, which is reduced to $\approx 10\%$ for the 
larger $Q^2$. 

HE3): $\,\,R_{L,T}$ for approximately constant $|\bmq|= $1.05 and running $\nu$ 
\cite{chen1,chen2}. 

Eq. (\ref{a13b}), using $R_L$ is a simpler expression than Eq. (\ref{a13a}),
for reduced, total inclusive cross sections and requires only  additional
knowledge of $\alpha_p$. Extracted $\alpha_n$ from, in principle favored $R_T$
data, ought to be close to the ones from cross section data HE2) with 
approximately the same $Q^2$, yet Table I shows fairly large deviations. 
The following may well be one of the causes.

In order to be eligible as partners for a Rosenbluth extraction, some data 
sets HE2) have been chosen for fixed $\theta=85^{\circ}$ and at two beam 
energies, such that the $x,Q^2$ $approximately$ coincide with those of the 
first set with $\theta=15^{\circ}$ at different beam energies. Since such 
a match is never perfect, extrapolations of data are necessary. 

$R_T$ is relatively large for $x\approx 1$, but that is also region where 
the point-nucleon SF $f^{PN,A}(x,Q^2)$ varies sharply with $x$. A 3$\%$
error or uncertainty in an extrapolation of data points to values for 
$x \approx 1$, may cause a $10\%$ change in $f^{PN,A}$ on which $F_2^{A,NE}$ 
depends linearly. We thus tend to actually  trust more the involved total 
cross section information. A PWIA analysis of $R_T$ is reported 
to be in good agreement with data on the elastic side of the QEP \cite{ciofi}.

The following D data sets are nearly all for appreciably larger $Q^2$ than 
for He.

D1): \,\,$E=4.045$ GeV; $\theta=15,23,30,37,45,55^{\circ}$;
${\bar Q}^2=0.972, 1.940, 2.774, 3.535, 4.251,4 .900$ GeV$^2$ 
\cite{nicu1,arrd}.

In our previous analysis we searched for cross section data dominated by 
$F_2^p$. This limited the suitable data to the two lowest $\theta$ sets 
above~\cite{rtv}. With reliable parameterizations for both $F_{1,2}^p$, 
the above restriction is lifted, providing a far larger number of data 
points for analysis.

Only in the data for the lowest two angles is a pseudo-resonance clearly
visible: For $\theta \ge 30^{\circ}$ the pseudo-resonance structure
gradually disappears. One still notices a minimum beyond the QEP, but
for further growing $Q^2$, there remains no more than a break in the slope
of ln($\,\sigma$). The agreement over the entire range of $\nu$, with 
the exception of the intermediate range, discussed at the end of Section 
IV, is good and sometimes excellent. It comprises qualitatively different 
features, as are the position and intensity at the minimum between 
the QEP and the pseudo-resonance, as well as the position of the peak of 
the pseudo-resonance. In those regions the NI components first compete with 
NE, overtake those and finally dominate. 

An underestimate of the computed NI in the intermediate $\nu$-range, and its
remarkably similar relative size is apparent in virtually all analyses. For 
example, Figs. 5, 2 and 6 for $\theta=15, 30, 55^{\circ}$ show a bell-shaped 
excess over NE(extr) on the inelastic side of the QEP, as extracted from 
Eqs. (\ref{a1}), (\ref{a5a}), (\ref{a5b}). Dot-dashes show NI(emp), 
which produce NE($\langle \alpha_n \rangle$) for all $\nu$.

D2):  \,\,$E=5.507$ GeV, $\theta=15.15, 18.98, 22.81, 26.82^{\circ}$;
${\bar Q}^2=1.75, 2.50, 3.25, 4.00$ GeV$^2$ \cite{lung,stuart}.

The kinematics of the above NE11 data partly overlap those of D1). Virtually 
all remarks on D1) hold also for D2) (cf. Figs. 7,8 for $\theta=
15.15,26.81^{\circ}$.)

D3): \,\,$R_T$ for the $x,Q^2$ kinematics of D2 \cite{lung}. 

For kinematics close to those in D2), the extracted  $\alpha_n$ from $R_T$ 
are comparable or slightly higher than from D2). Otherwise much the same 
remarks as for HE3) above hold also here: In order to obtain $x,Q^2$ 
matching for a Rosenbluth separation, the unavoidable handling of data 
around the QEP  $x\approx 1$ is bound to be imprecise in view of the 
sensitivity of $f^{PN,D}$ around the QEP, which grows with $Q^2$. 

D4): \,\,$E=2.015$ GeV, $\theta=38.84^{\circ}$; $E=3.8$ GeV, 
$\theta=47.86^{\circ}$;  
$E=4.212$ GeV, $\theta=53.39^{\circ}$; $E=5.12$ GeV, $\theta=56.64^{\circ}$.  
${\bar Q}^2= 1.22, 3.15, 5.10, 6.83\,$ GeV$^2$ \cite{arrdx1}  

The above NE18 SLAC data for D are for high $E$ and relatively large
$\theta$ and are restricted to the immediate QE region. In spite of 
considerable experimental noise, those data are of interest, in view of 
the large $Q^2$ involved. Computed results are in agreement with data around 
the QEP, which show much scatter. Beyond that region, disagreements are 
less than $\approx 25\%$ (cf. Fig. 9). For reasons already mentioned we 
did not analyze parallel data for heavier targets. 
  
D5): \,\,$E= 9.744, 12.565, 15.730, 17.301, 18.476, 20.999\,$ GeV, $\theta=
10^{\circ}.\,$ ${\bar Q}^2= 2.5, 4.0, 6.0, 7.1, 8.0, 10.0\,$ GeV $^2$ 
\cite{rock}.

Cross sections computed with rather primitive input show on a tight 
logarithmic scale reasonable agreement with the above old data, but on a 
linear scale considerable scatter in data is apparent.  The above Rock 
data for small $\theta$ provide a unique example of marginal stability:
the $E=9.744$ provide two adjacent subsets with rather different average
for $\alpha_n$. Both cause $\langle \alpha_n \rangle $ to have relatively 
large error bars (Table I). 

The main interest 
is their values out to the largest $Q^2$, measured until this date. 
Figs. 10-12 show results for $E=15.73, 18.476, 20.999\,$ GeV.
Table I shows that for $E=18.476$ $\,\,\langle \alpha_n \rangle$ 
is relatively large, and not in line with other $\langle \alpha_n\rangle$ 
in the data set.  One observes close correspondence between 
$\langle \alpha_n \rangle$, extracted from different data sets.

The overall outcome is compiled in Figs. 13-15. In inserts we give symbols
for experiments (empty ones for older and filled symbols for re-analyzed
ones), kind if data and Ref. numbers. Fig. 13 contains results from He 
data, which are all for low $Q^2\le 1.3\,$GeV$^2$ with $\alpha_n$, close 
to 1. For those $Q^2$, differences for the choices I, II are relatively small 
(see Table I); displayed results are for I. In detail:

i) Asymmetry measurements in QE $\vec{^3{\rm He}}(\vec{e},e')$ 
\cite{gao,xu,xu1}. 

ii) Preliminary results of the analysis of i), without 3-body channels 
in the FSI \cite{kps}.  

iii) Ratios of exclusive D break-up cross sections, with $p$, respectively 
$n$ detected \cite{anklin1,anklin2,kubon} (In view of the criticism of 
Jourdan $et\,al$. \cite{jourd}, we omit some old data \cite{bruins}).

iv) The above HE1)-HE3) \cite{ne3,chen1,chen2} $\vec{^3{\rm He}}
(\vec{e},e')$.

Fig. 14 contains results, extracted  from the sets D1)-D5) with $Q^2\ge 1.0\,
$GeV$^2$ \cite{nicu1,arrd,stuart,lung,arrdx1,rock}. The only displayed 
result of a previous analysis is D2) \cite{lung}. Fig. 15 is the same as 
Fig. 14 for choice II of FFs.  

In both the drawn lines are inverse polynomial fits of $\alpha_n({\bar Q}^2)$ 
from the D data, as used in Ref. \cite{bba}
\begin{eqnarray}
{\alpha_n({\bar Q}^2)}^I   &=& 1/[1+0.007437 {\bar Q}^2 +0.002815161 
{\bar Q}^4 -0.000115008 {\bar Q}^6]
\nonumber\\
{\alpha_n({\bar Q}^2)}^{II}&=& 1/[1+0.06568 {\bar Q}^2 -0.00678662 {\bar Q}^4+ 
0.000323355 {\bar Q}^6]
\label{a14}
\end{eqnarray}
Although only approximately valid for ${\bar Q}^2\ge 1.5\,{\rm GeV}^2)$,
the fits extrapolate to the $'$correct$'$ $\alpha_n(0)=1$.  In spite 
of scatter in the data two observations on $\langle \alpha_n \rangle$  
stand out:\\
a) $\alpha_n({\bar Q^2})$ clearly decreases with increasing $Q^2$\\  
b) There is close correspondence between extracted values from different 
data sets,  for instance for ${\bar Q}^2=2.5, 4.0,\approx 7.0\,$GeV$^2$.

\section{Summary and Conclusions.}

The overwhelming majority of its properties  cannot been measured on free 
neutrons and one is therefore led to study those on neutrons, which are bound 
in nuclear targets. The extraction of any such property thus demands an 
accurate treatment of the embedding of the neutron in that target. Even when 
feasible, one has in addition to assume, that those extracted quantities 
are the same as for a neutron in vacuum, i.e. that the former are not 
$intrinsically$ modified by the medium. Nowadays one can accurately 
compute nuclear properties for $A\le 4$, and there seems to be no evidence 
for medium effects. Those may well be artifacts of approximations.

The present contribution deals with the extraction of the reduced 
magnetic form factor of the neutron from cross sections for the inclusive 
scattering of electrons from the lightest targets. 

Part of the available data have been analyzed before ~\cite{rtv}. There 
we limited ourselves to those kinematic parts of data sets, where the
elastic absorption of an exchanged photon on a nucleon had been estimated 
to be much in excess of the inelastic ones, which were subsequently 
disregarded: That procedure not only affected accuracy. It also limited 
the analysis to the immediate neighborhood of QE peaks, where the NI 
background is very small. That region covers only a small section of the 
data, which generally stretch over wide energy-loss ranges.

Ref. \cite{rtv} followed the principal ideas of Lung \cite{lung} and of 
older work, using rather primitive tools and input\cite{rock}. A number of 
incentives called for a re-analysis of the above results. Together with 
our wish to include the above mentioned neglected kinematic areas, we 
included previously disregarded data. We  used moreover recently published, 
precise parameterizations of the input SF $F_{1,2}^p$ for the proton, 
valid through resonances and reaching into the DIS region. For medium and 
large $Q^2$ those parametrizations become unreliable and  we had to 
fall back on parameterizations of resonance-averaged $F_2^p$. In addition,  
it appeared possible to up-date the input for the determination of the 
additionally required neutron SF. Finally we could also sharpen some 
theoretical tools. As a result the present analysis constitutes a 
significant refinement of the previous one.

The above comprehensive theory for the extraction of the magnetic form 
factor of the neutron $G_M^n$ addressed relatively abundant total 
cross for inclusive electron scattering, and scarce data on their 
transverse components. In principle several targets are accessible to an 
analysis, but only for the lightest nuclei can one presently calculate 
with great precision  nuclear information, which describes the above 
embedding.  Most of the available data on D are of good quality and 
contain several data sets covering a range of partly overlapping $Q^2$. 
The latter fact enables desirable consistency checks. For $^4$He 
there are only available rather old, low-$Q^2$ data of lesser quality
and for which also the theory is less accurate than for higher $Q^2$.
Nevertheless we analyzed all.

The cornerstone of our analysis is the possibility to reliably compute 
components of the nuclear inclusive cross sections, due to inelastic
virtual photon absorption on nucleons, provided  $F_k^N$ are known.
Those nuclear NI processes dominate virtually all kinematic regions, 
except the ones around the QE peak, where elastic absorption of virtual 
photons on nucleons competes with inelastic processes.
 
Cross sections in those QE regions contain the desired information on 
form factors, and their isolation is therefore of primary importance. 
Simple theoretical considerations predicts the same $x$-dependence of 
the NE components of cross sections and the calculable Structure Function 
$f^{PN,A}(x,Q^2)$ of a fictitious target, composed of point-nucleons. The 
latter drops sharply from its QE peak at $x\approx 1$: in the case of a D 
by a factor $\approx 10-50$. 

One thus compares the functional dependence on $x$ of the above difference 
with the same for $f^{PN,A}(x,Q^2)$. In regions where close similarity is
found, one identifies that difference with the desired NE components of
cross sections.  

For all data sets, the above differences between measured reduced total
cross sections and calculable inelastic backgrounds, appear to follow the 
predicted $x$-behaviour, roughly for $x \gtrsim 0.7-0.8$. The above provides 
incontrovertible proof that in the above restricted areas, where the 
$x$-dependence is most outspoken, the above defined differences are indeed  
the NE components of the cross sections. 

For each data set we then selected a continuous $x$-range, for which the 
correspondence with the $x$-dependence of $f^{PN,A}$ is best. Dependent on the 
quality of the data set, the number of thus selected $x$ points may be as 
large as 17. Ideally, those should reproduce $G_M^n$, independent of $x$ or 
$A$.

We first summarize our results: 

A) There is  similarity and sometimes reasonable correspondence between
the extracted $\alpha_n$ from different targets. Differences may in part 
be due to the same in the quality of the older low $Q^2$, He data and 
the more recent D experiments. It is in particular satisfactory that 
the high-quality $E=4.045$ and $E=5.507$ D data yield corresponding 
$\alpha_n(Q^2)$ for similar $Q^2$. The same is the case, if the old 
data of Rock $et\,al.$ are included.
 
B) Our results confirm and reinforce an important conclusion already
reached in our previous analysis: $\alpha_n$ from D data seems to fall 
faster with $Q^2$ than the older Lung results indicate. The same is the 
case for the slope of the older He data, but as stated above, we have
reservations regarding the used parametrization for $F_2^p$ for the 
lowest $Q^2$ and the quality of the data. For growing $Q^2$, the NE3 
and NE9 $^4$He data and the average behavior of the D data for 
corresponding $Q^2$ produce  similar $\alpha_n$.

We have emphasized the identification of $\sigma^{A,NE}$ over finite
$x$-ranges. On the inelastic side of the QEP for virtually all D data sets, 
the use of NI, computed with the recent parametrization of $F_2^p$, produces  
deviations from NE components, computed with $\langle \alpha_n \rangle$. 
Those roughly start where NE$\approx$NI, and grow to a 10-15 $\%$
under-estimate  towards the peak of the pseudo-resonance position, beyond
which the discrepancy rapidly disappears. An increase of the NI background
of just the above size extends the above local fit of the computed NE cross 
section to comprise the entire $\nu$-range.

The discrepancy may be the result of a relatively modest under-estimate 
of the transition strength for inclusive excitation of the first resonance 
for larger $Q^2$. It would effect mostly the low $\nu$ tail of NI, because 
more inelastic parts are screened by overlapping resonances. We emphasized
that one cannot apply an ad hoc change of one parameter in a multi-parameter
fit of $F_2^p$.

In spite of the manifestly succesful local isolation of elastic 
components of the inclusive cross sections, the actual extraction 
of $G_M^n$ from those is not straightforward. We mention a few sources 
which may influence the precision of the extraction:

1) Disregarding experimental accuracies, the test of the central 
requirement is  optimal when the inelastic background is a small fraction 
of the total cross section. The latter is the case in the immediate
neighborhood of the QEP, where the elastic components are most sensitive to
variations of $G_M^n$. The small values of the inelastic background there
are due to the tails of overlapping resonances, which by nature have least 
accuracy. Consequently, changes in a very small background around the QEP
may have $1\%$ effects on the extracted $\alpha_n$.

2) Total inclusive cross sections contain in principle Meson Exchange 
Contributions (MEC), which should be eliminated from the data, before
those are manipulated as described. Little is known of MEC contributions for 
high-$Q^2$ inclusive processes. As a measure of their size we suggested 
recent results for the exclusive processes $^3$He(e,e'p)D and 
$^4$He(e,e'p)$^3$He. Polarization variables appear sizeably affected by MEC, 
but only tiny corrections are reported on cross sections for inclusive 
scattering with unpolarized beams \cite{rocco1}.  

3) Present uncertainties regarding the proton form factors.

The above concludes our program to extract $G_M^n$ from data on inclusive
cross sections. We are well aware that our method is an indirect one, which
forced us to make careful checks on intermediate steps. Those unfortunately 
could not circumvent present uncertainties in input. At least the unknown 
behavior of $G_E^n$ for large $Q^2$ seems not to be of relevance. Unless 
the unknown $G_E^n$ for larger $Q^2$ will turn out to  deviate strongly from 
assumed extrapolations from low $Q^2$, its influence will remain negligible. 
         
The above clearly demonstrates that the described extraction method is  
a realistic and internally consistent one. We do not think that theoretical 
tools can be much improved, but it is highly desirable to eliminate 
uncertainties in some input elements of the calculations. In parallel new 
inclusive scattering data  on $^3$He and $^4$He could add information 
and furnish further proof of consistency. 

In the meantime we are looking forward to final results  for $G_M^n$
from the D$(e,e'p)X$/$D(e,e'n)X$ experiment of Brooks $et\, al$. 
Preliminary results seem to yield a smaller slope for $\alpha_n$ than
in Figs. 14,15 \cite{brooks}.

\section{Acknowledgments.}

ASR is much indebted to Cynthia Keppel, Eric Christy and John Arrington 
for generously making available parameterizations of proton structure
functions, prior to publication and J.A. for having added relevant comments.
He is grateful to J-P. Chen, who provided unpublished experimental 
material, simultaneously making persistent and correct critical remarks on 
the role of the NI background.

\begin{figure}[p]
\includegraphics[scale=.8]{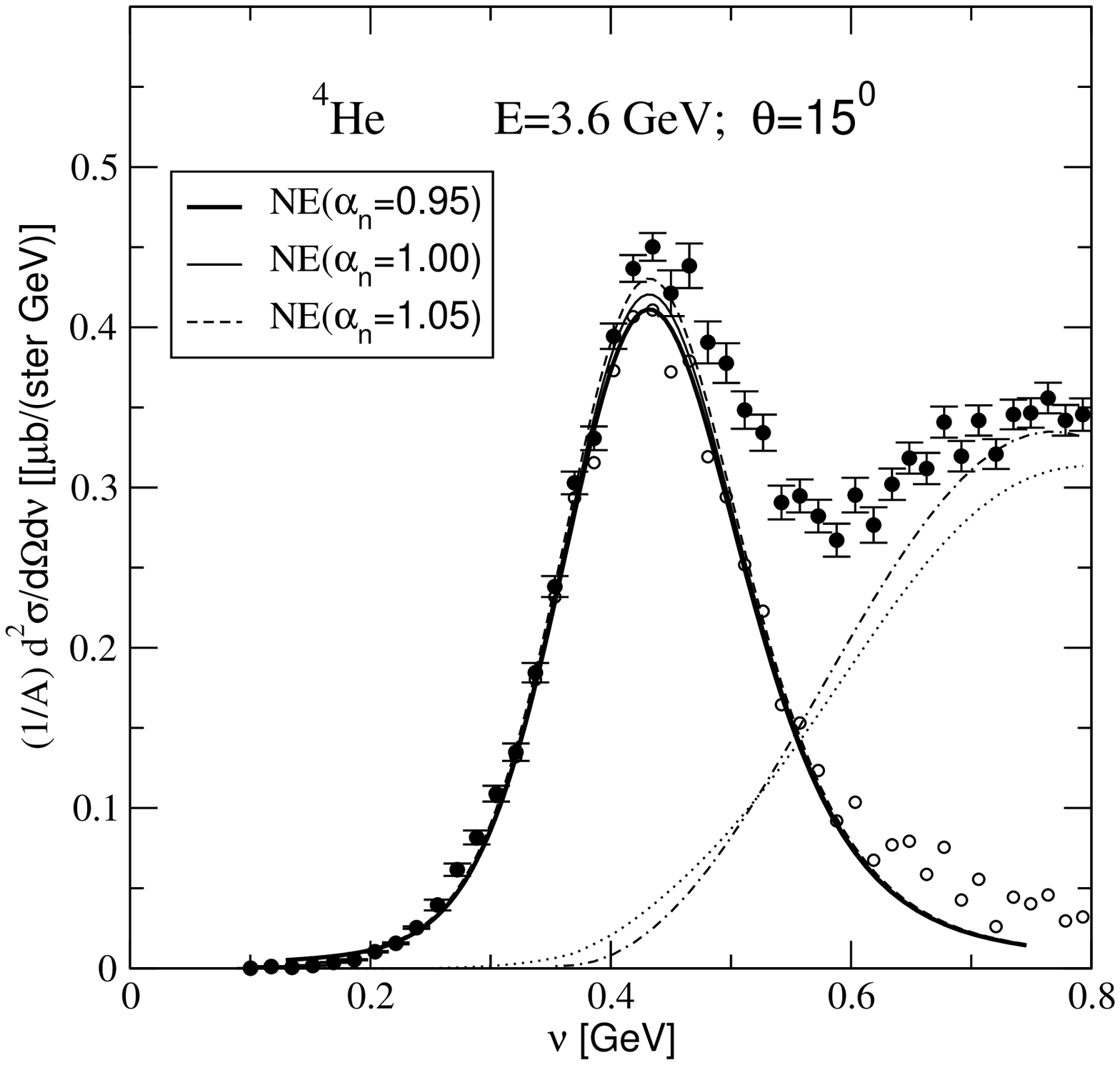}
\caption{Inclusive cross section on $^4$He for $E=3.6$ GeV, $\theta=
15^{\circ}$, ${\bar Q}^2=0.781$ GeV$^2$. Filled and empty circles represent 
SLAC NE9 data~\protect\cite{chen2} with their statistical errors, 
respectively cross sections, from which the $theoretical$ NI part (dotted 
curve) has been subtracted. Empty circles should carry error bars coming from
the data and the NI part, but since we cannot at present estimate those,we
have preferred not show them.
up to the last extraction point used. The difference is compared with 
$\sigma^{{\rm He},NE}$ for $\alpha_n=1.00, 1.00\pm 0.05$ (curves). Dot- 
dashes represent an NI(emp), defined as $\sigma^{tot}-\sigma^{NE}$.}
\end{figure}

\begin{figure}[p]
\includegraphics[scale=.8]{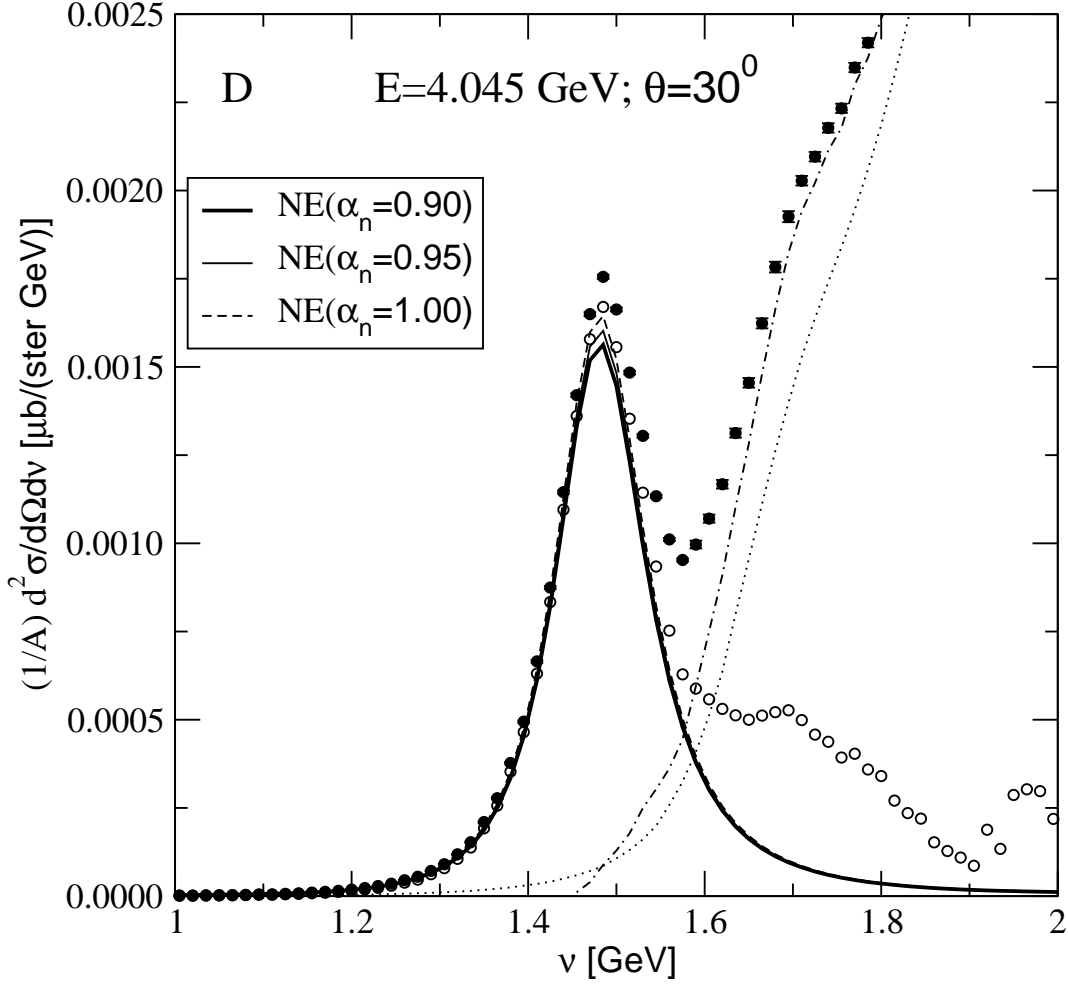}
\caption{As Fig. 1 for D at $E=4.045 {\rm GeV},\theta=30^{\circ}$,
${\bar Q}^2=2.774\,$GeV$^2$. Data:  Jlab E89-009~\protect\cite{nicu1}. NE 
curves are for $\alpha_n=0.95, 0.95\pm 0.05$.}
\end{figure}

\begin{figure}[p]
\includegraphics[scale=.8]{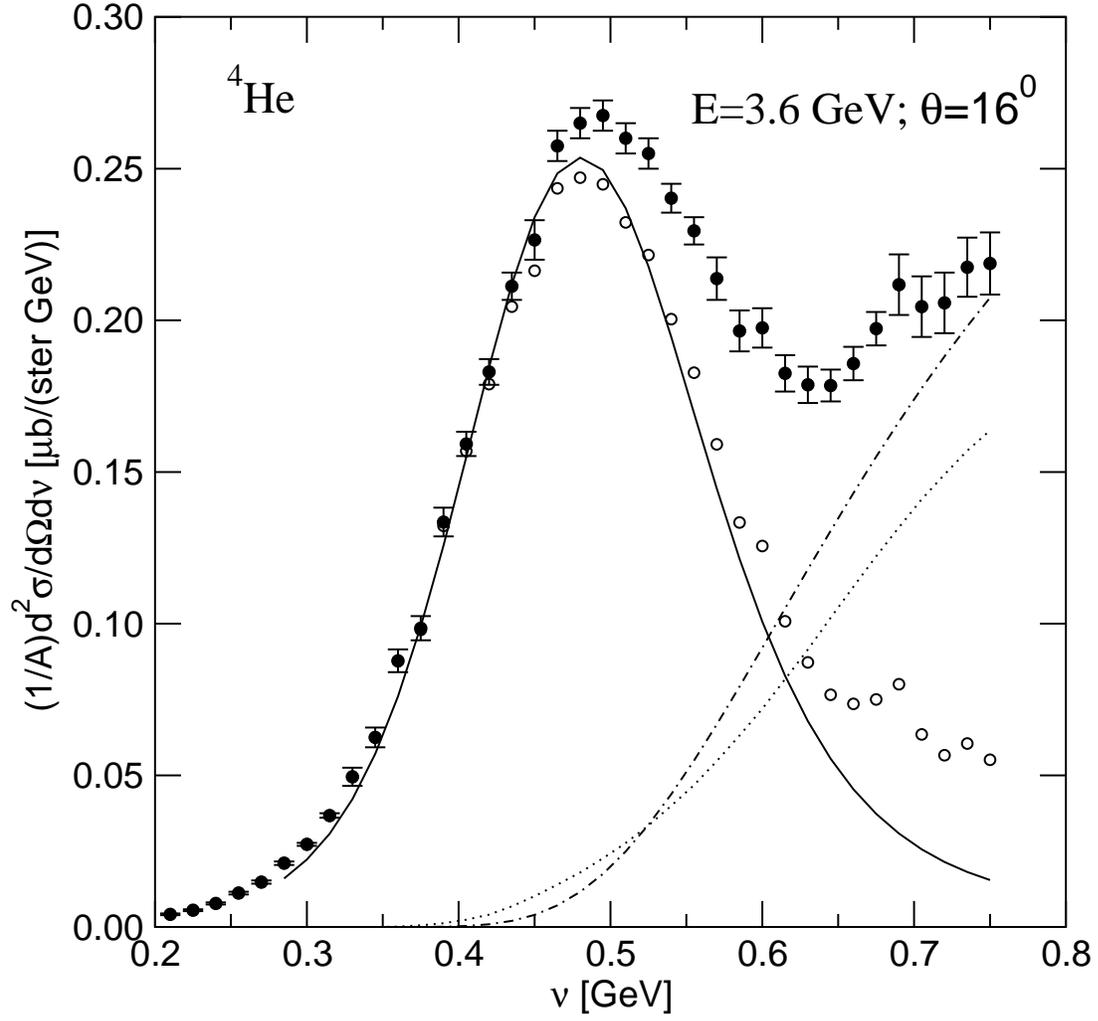}
\caption{As Fig. 1 for $^4$He at $E=3.595$ GeV, $\theta=16^{\circ}$. 
${\bar Q}^2=0.873\,$GeV$^2$. Data: SLAC-Virginia NE3~\protect\cite{ne3}. 
NE curve for $\alpha_n=1.006$. }
\end{figure}

\begin{figure}[p]
\includegraphics[scale=.8]{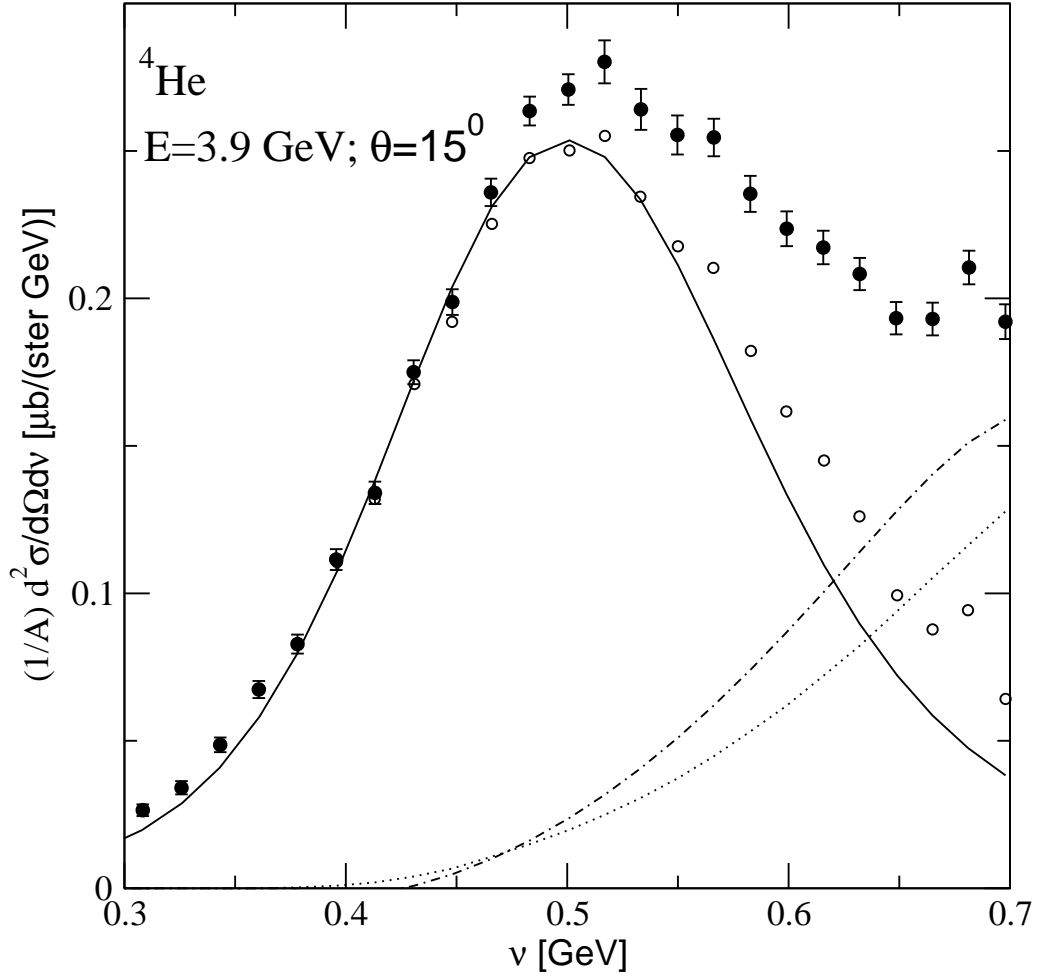}
\caption{As Fig. 1 for $^4$He at  $E=3.9$ GeV, $\theta= 15^{\circ}$,
${\bar Q}^2=0.907\,$GeV$^2$. Data: SLAC NE9~\protect\cite{chen2}. NE curve 
for  $\alpha_n=1.007$. }
\end{figure}

\begin{figure}[p]
\includegraphics[scale=.8]{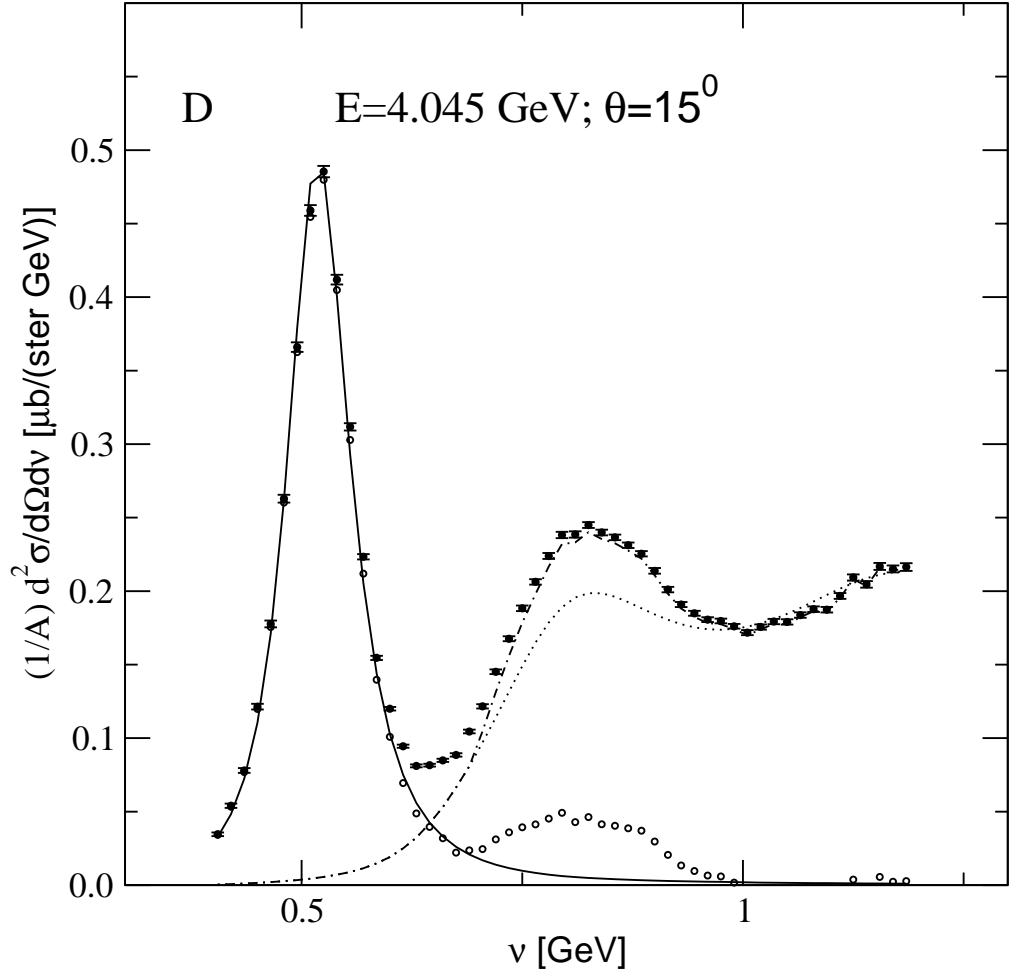}
\caption{As Fig. 1 for D at  $E=4.045$ GeV, $\theta= 15^{\circ}$,
${\bar Q}^2=0.972\,$GeV$^2$. Data: JLab E89-009 ~\protect\cite{nicu1,arrd}. 
NE curve for $\alpha_n=$1.004. }
\end{figure}

\begin{figure}[p]
\includegraphics[scale=.8]{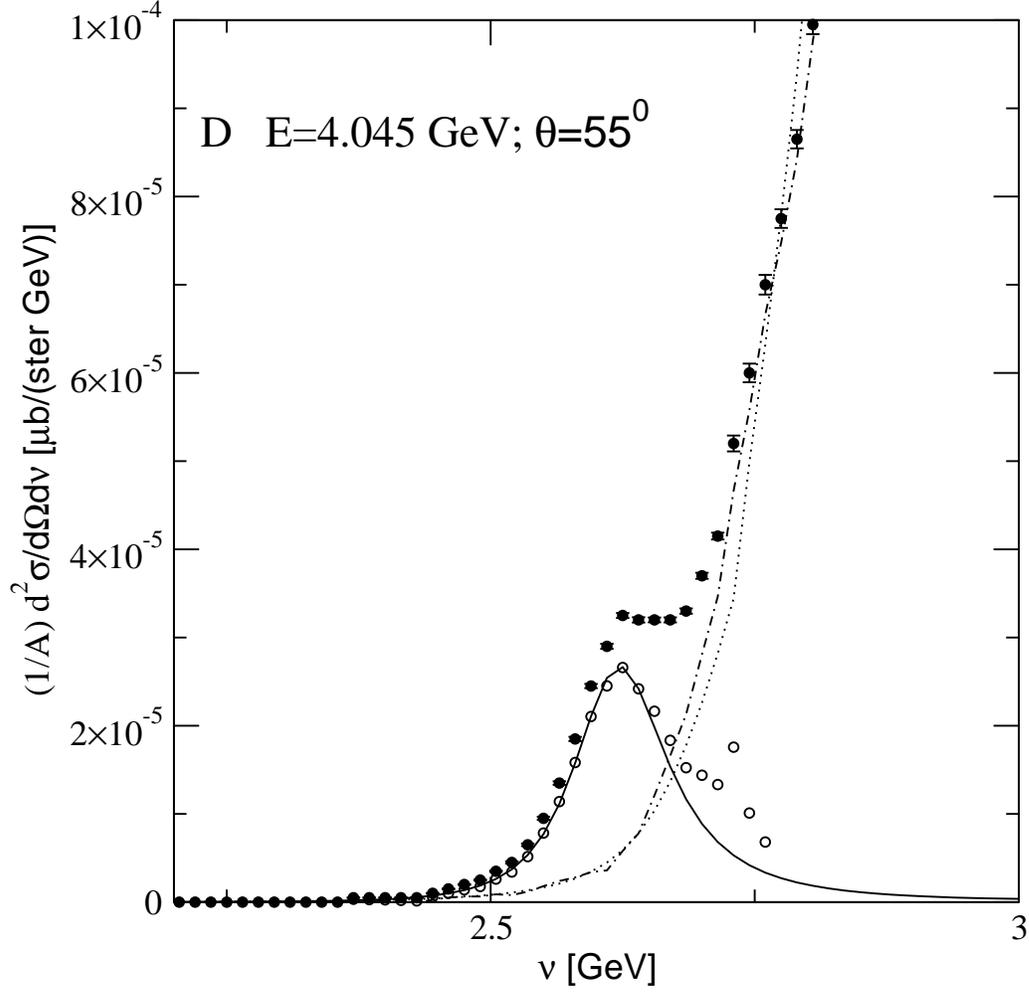}
\caption{As Fig. 1 for D at  $E=4.045$ GeV, $\theta= 55^{\circ}$,
${\bar Q}^2=4.900\,$GeV$^2$. Data: JLab~\protect\cite{nicu1}. NE curve for 
$\alpha_n=0.991$. }
\end{figure}

\begin{figure}[p]
\includegraphics[scale=.8]{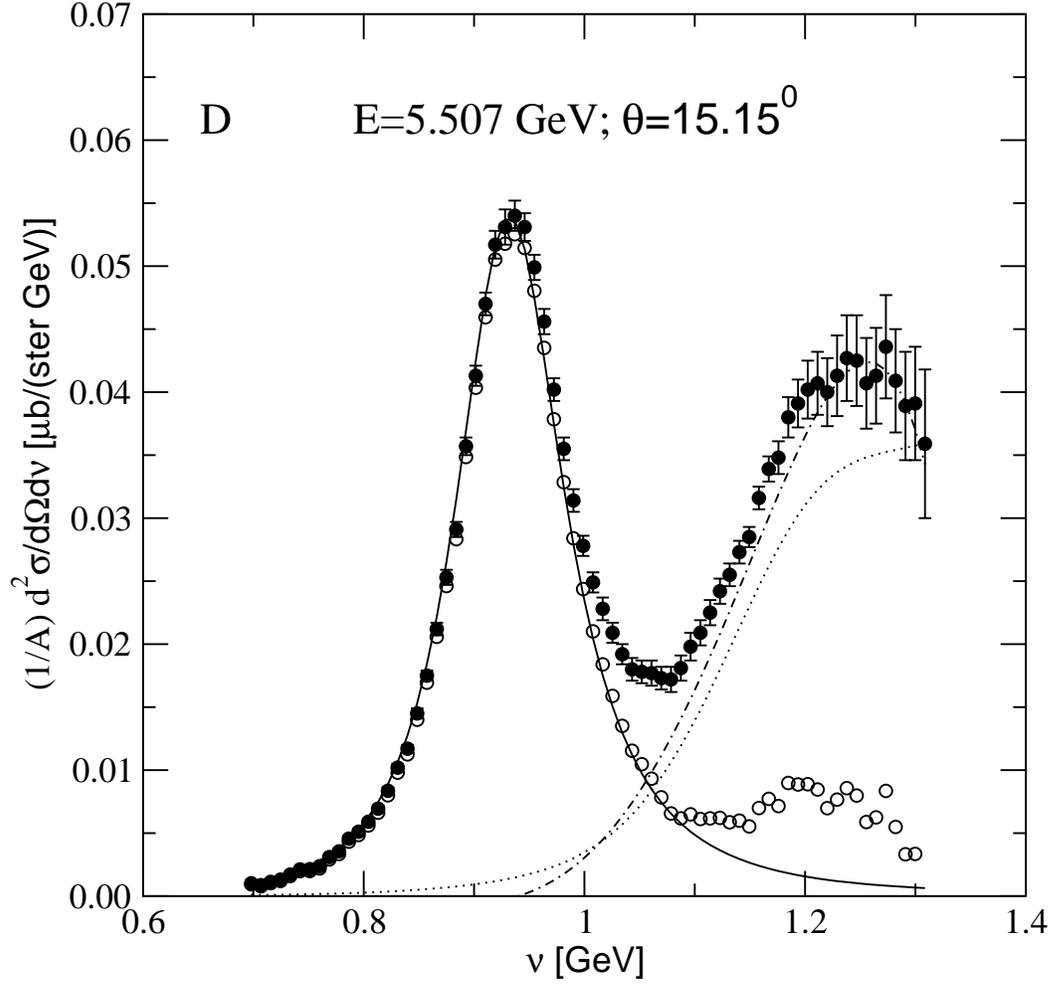}
\caption{As Fig. 1 for D at  $E=5.507$ GeV, $\theta= 15.15^{\circ}$,
${\bar Q}^2=2.750\,$GeV$^2$. Data: SLAC NE11~\protect\cite{stuart}. NE curve 
for $\alpha_n=1.008$. }
\end{figure}

\begin{figure}[p]
\includegraphics[scale=.8]{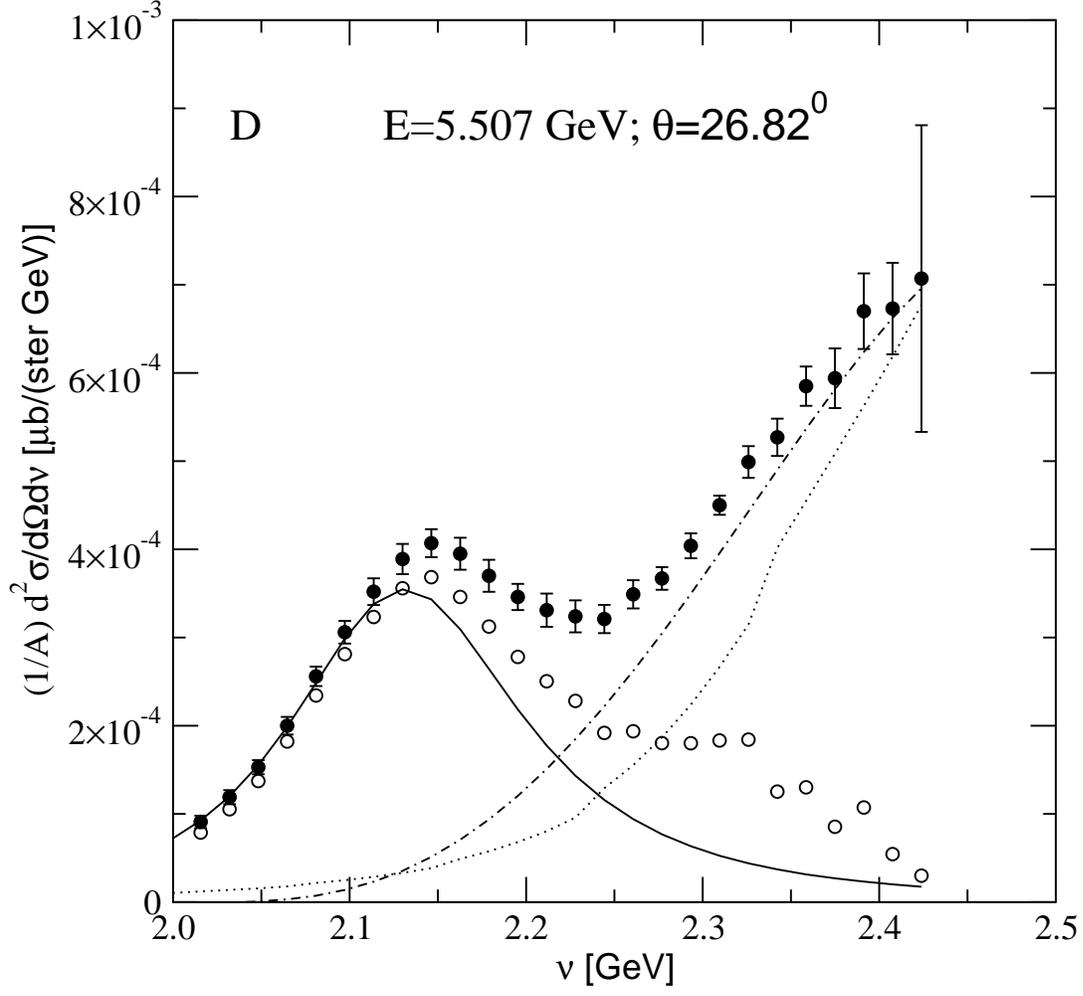}
\caption{As Fig. 1 for D at  $E=5.507$ GeV, $\theta= 26.86^{\circ}$,
${\bar Q}^2=4.000\,$GeV$^2$. Data: SLAC NE11~\protect\cite{stuart}. NE curve 
for $\alpha_n=0.951$. }
\end{figure}

\begin{figure}[p]
\includegraphics[scale=.8]{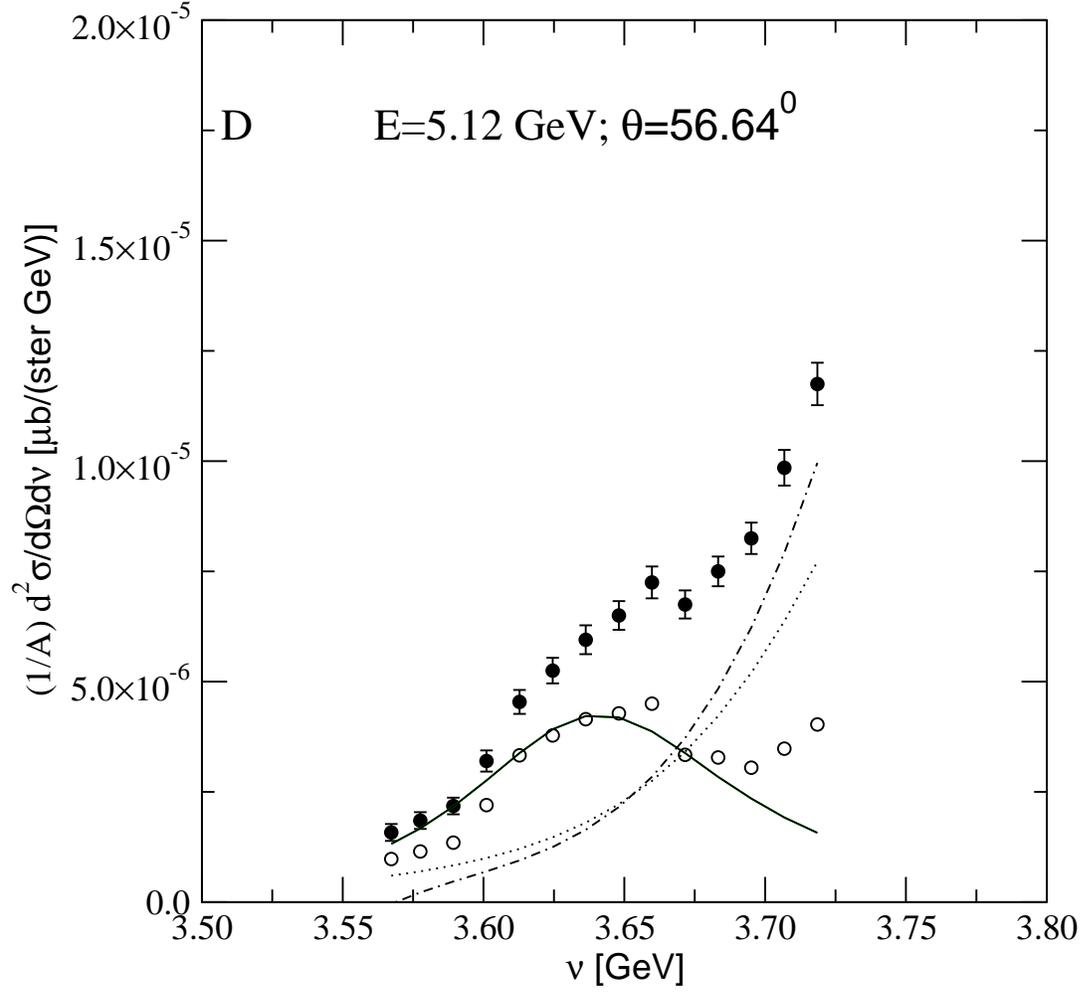}
\caption{As Fig. 1 for D at  $E=5.12$ GeV, $\theta= 56.64^{\circ}$,
${\bar Q}^2=6.83\,$GeV$^2$. Data: SLAC NE18~\protect\cite{arrd}. NE curve 
for $\alpha_n=0.842$. }
\end{figure}

\begin{figure}[p]
\includegraphics[scale=.8]{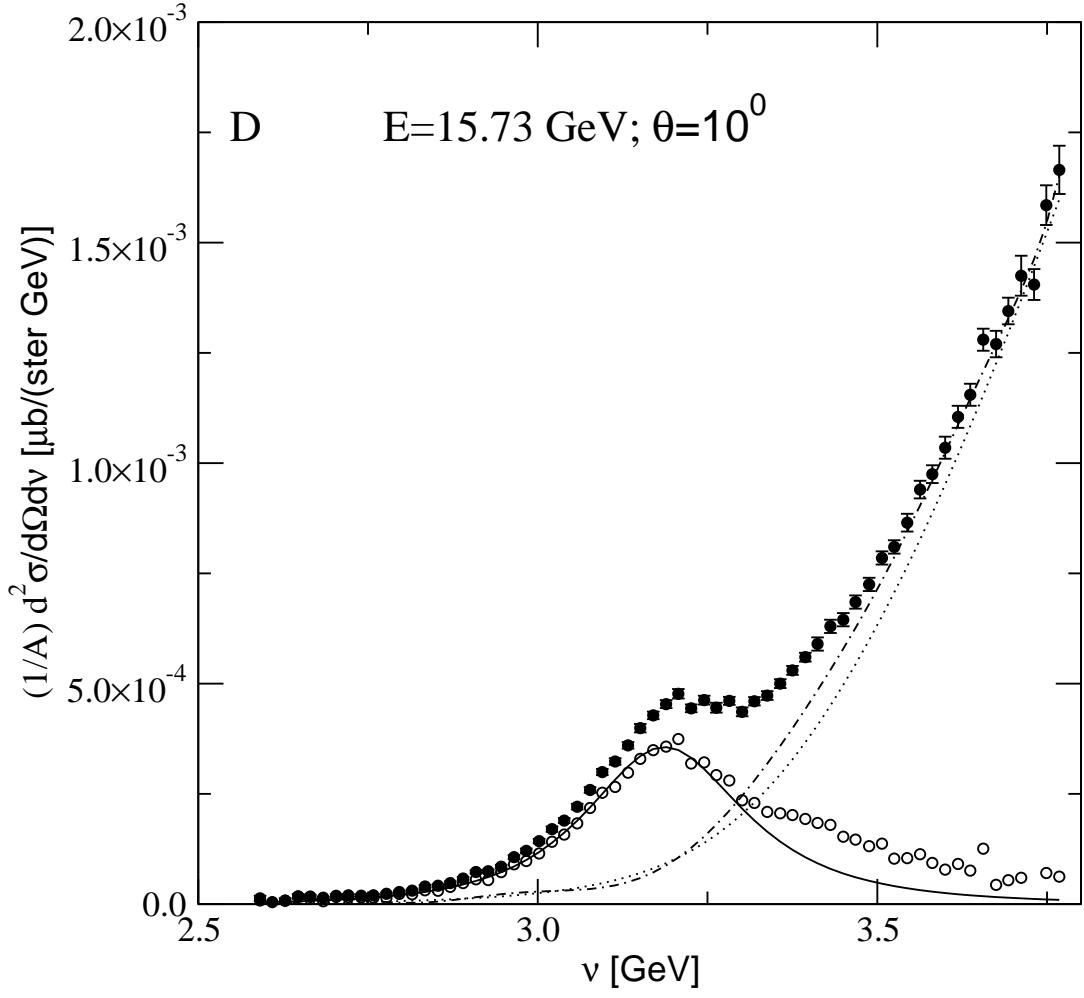}
\caption{As Fig. 1 for D at  $E=15.73\,$ GeV, $\theta= 10^{\circ}$, 
${\bar Q}^2=6.00\,$GeV$^2$. SLAC E133 data ~\protect\cite{rock}. NE curve 
for $\alpha_n=0.854$ .}
\end{figure}

\begin{figure}[p]
\includegraphics[scale=.8]{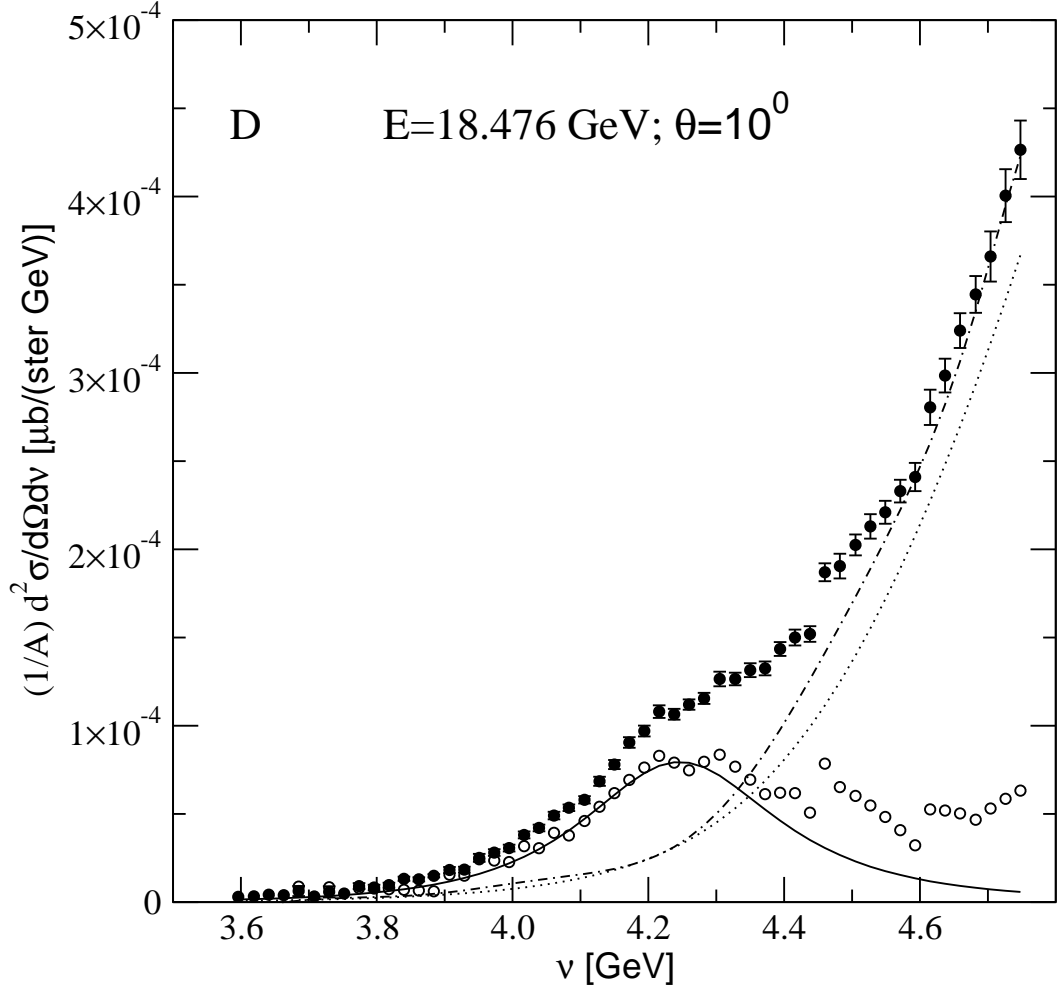}
\caption{As Fig. 1 for D at  $E=17.301$ GeV, $\theta= 10^{\circ}$.
${\bar Q}^2=7.12\,$GeV$^2$. SLAC E133 data ~\protect\cite{rock}. NE curve  
for $\alpha_n=0.879$ .}     
\end{figure}

\begin{figure}[p]
\includegraphics[scale=.8]{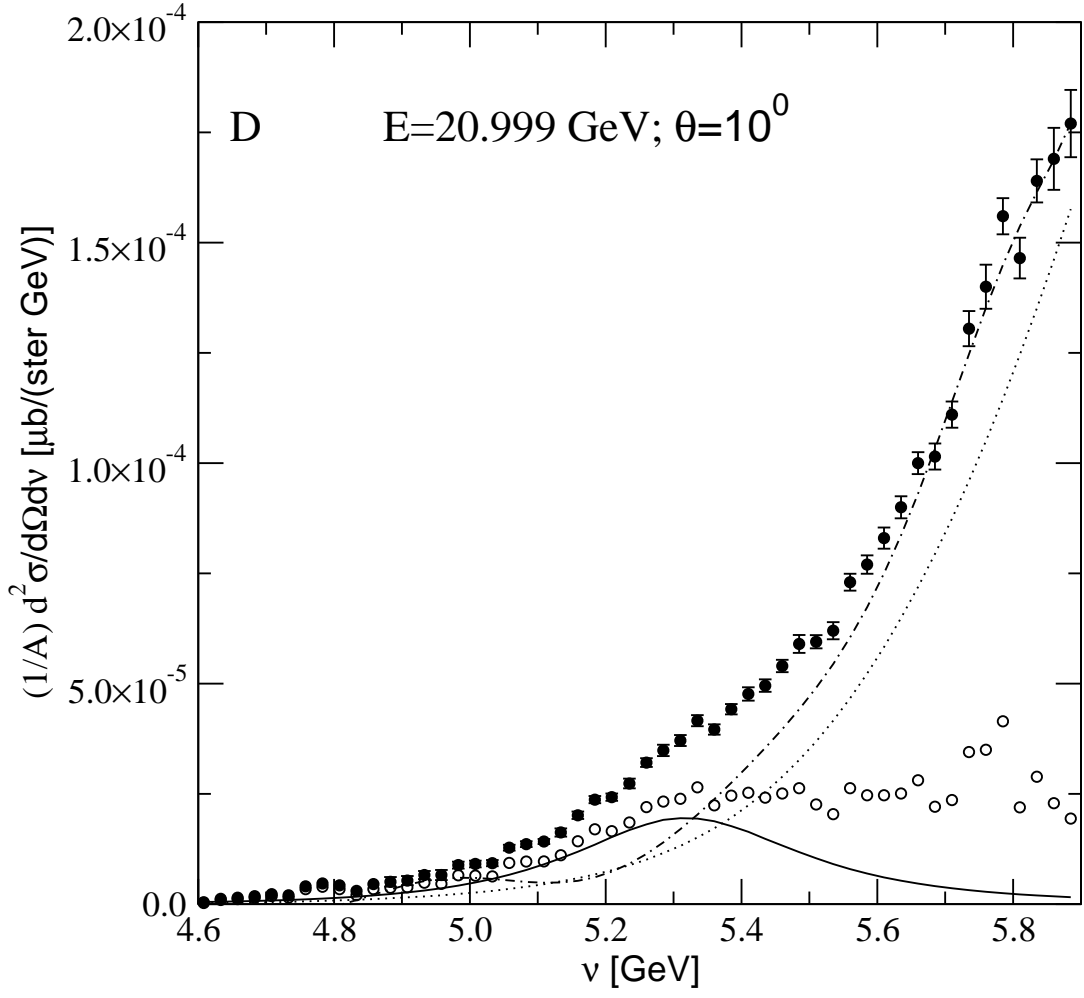}
\caption{As Fig. 1 for D at  $E=20.999$ GeV, $\theta= 10^{\circ}$.
${\bar Q}^2=10.06\,$GeV$^2$. SLAC E133 data ~\protect\cite{rock}. NE curve
for $\alpha_n=0.769$ .}
\end{figure}

\begin{figure}[p]
\includegraphics[bb=-150 440 567 400,angle=-90,scale=.65]{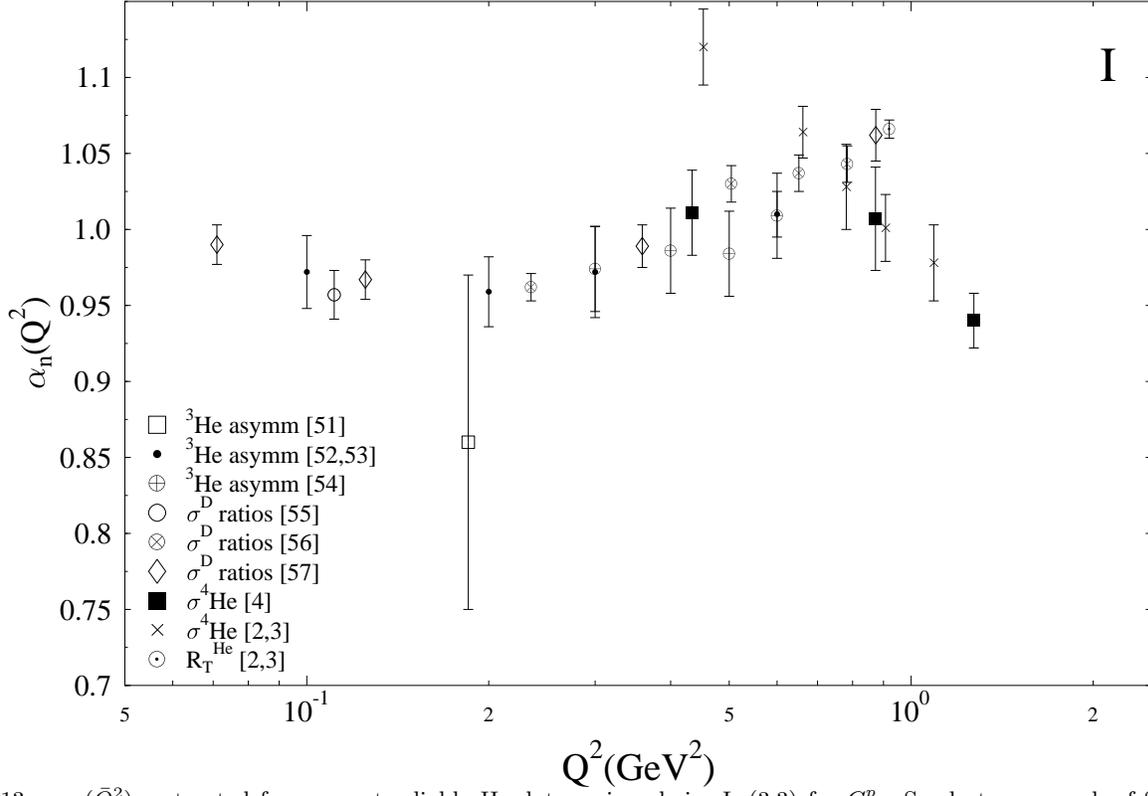}
\caption{ $\alpha_n({\bar Q}^2$), extracted from recent reliable He data, 
using choice I, (3.3) for $G_E^p$. See last paragraph of Section 
V for description.}
\end{figure}

\begin{figure}[p]
\includegraphics[bb=-150 440 567 400,angle=-90,scale=.65]{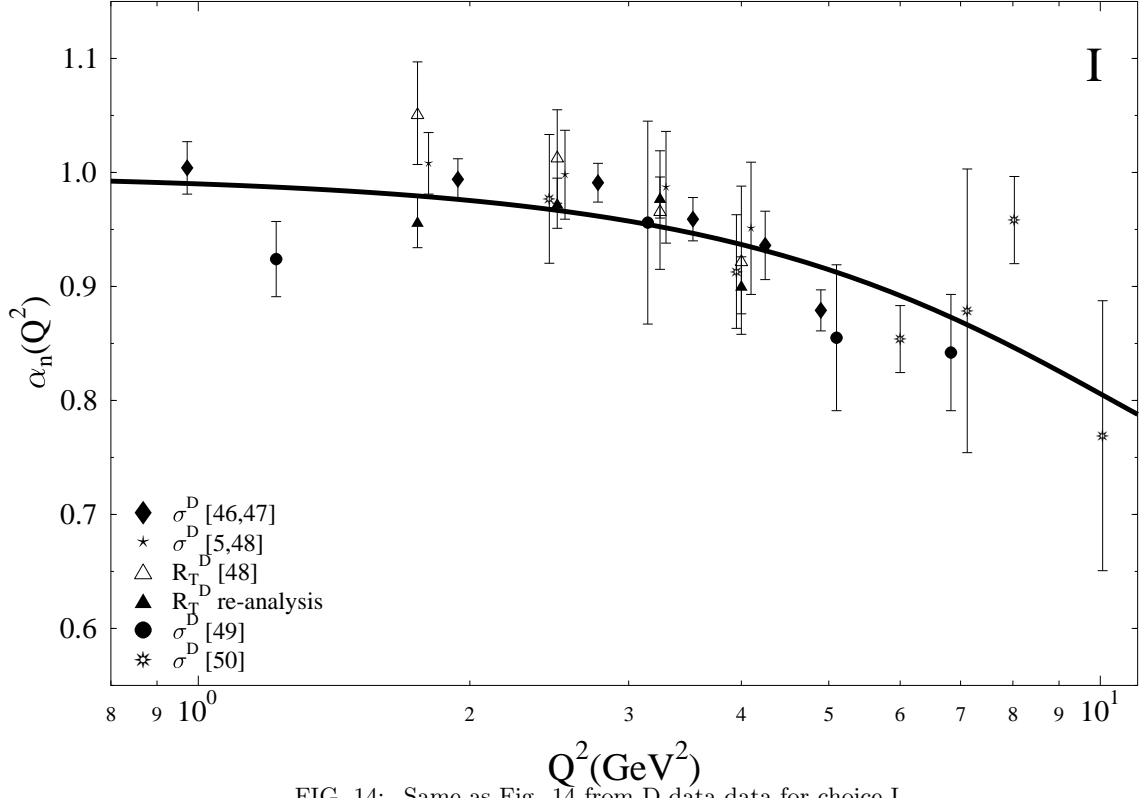}
\caption{ Same as Fig. 14 from D data data for choice I.}
\end{figure}

\begin{figure}[p]
\includegraphics[bb=-150 440 567 400,angle=-90,scale=.65]{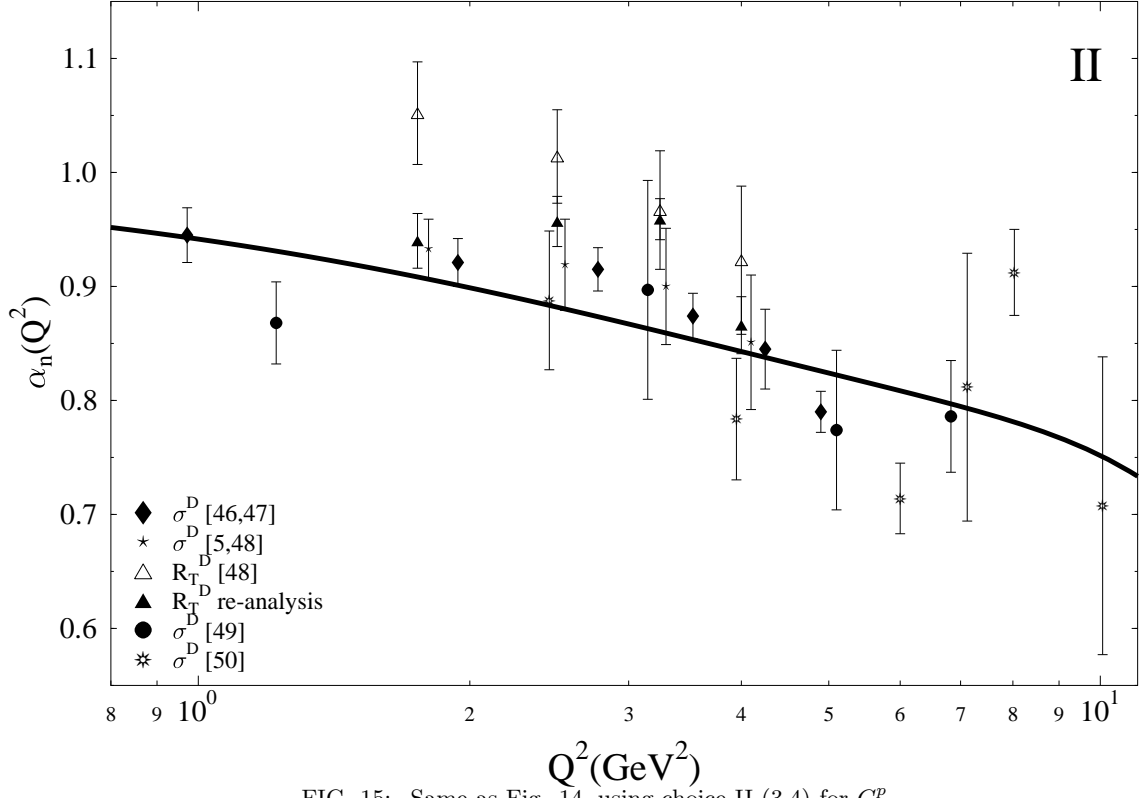}
\caption{ Same as Fig. 14, using choice II  (3.4) for $G_E^p$.}
\end{figure}

\newpage

\begin{table}
{\small 
\caption {Extraction of $\alpha_n(Q^2)$ from QE inclusive scattering data on
D, $^4$He. Columns 1-3 give the target, beam energy $E$ and scattering angle
$\theta$. Columns 4-6 indicate the ranges of $\nu,x,Q^2$, selected for the
analysis of the QE and pseudo-resonance region, and ${\bar Q}^2$ at the QEP.
The last column gives the number of points, selected for the extraction of
$\langle \alpha_n({\bar Q}^2) \rangle$, its average (with $Q^2\to{\bar Q}^2$
for identification and error of the mean for the new, respectively
the old FF parametrization.}

\begin{tabular}{c|c|c|c|c|cc|ccc|}
\hline
target  & $E$\ [GeV] & $\theta$  & $\nu$[GeV] & $x$&
$Q^2\ {\rm [GeV^2]};$ & ${\bar Q}^2$ & $n$
&[$\langle \alpha_n({\bar Q}^2)\rangle_O]; $
&[$\langle \alpha_n({\bar Q}^2)\rangle_N] $]\\
\hline
$^4$He$\,${\protect\cite{ne3}}
        & 2.02  & $20^{\circ}$ & 0.120-0.435 & 2.054-0.473&
          0.463-0.386; & 0.434 &  5    & 1.011$\pm 0.032;$ &1.008$\pm0.032$
\\
        & 3.595 & $16^{\circ}$ & 0.180-0.750 & 2.814-0.563&
          0.951-0.792; & 0.873 & 11    & 1.006$\pm 0.045;$ &0.954$\pm0.045$

\\
        & 3.595 & $20^{\circ}$ & 0.255-0.870 & 3.024-0.723&
          1.450-1.180; & 1.270 &  8    & 0.940$\pm 0.031;$ &0.865$\pm0.030$
\\
\hline
$^4$He$\,${\protect\cite{chen1}}
        & 2.7   & $15^{\circ}$ & 0.139-0.784  & 1.806-0.239&
          0.471-0.352; & 0.453 &  4    & 1.120$\pm 0.035;$ &1.114$\pm0.036$
\\
        & 3.3   & $15^{\circ}$ & 0.099-0.858  & 3.872-0.341&
          0.720-0.549; & 0.662 &  4    & 1.064$\pm 0.029;$ &1.031$\pm0.029$
\\
        & 3.6   & $15^{\circ}$ & 0.100-0.793  & 4.573-0.462&
          0.859-0.689; & 0.781 &  7    & 1.028$\pm 0.037;$ &0.982$\pm0.038$
\\
        & 3.9   & $15^{\circ}$ & 0.104-0.698  & 5.167-0.649&
          1.009-0.851; & 0.907 & 10    & 1.007$\pm 0.034;$ &0.950$\pm0.034$
\\
        & 4.3   & $15^{\circ}$ & 0.191-1.000  & 3.357-0.515&
          1.204-0.967; & 1.090 &  9    & 0.978$\pm 0.033;$ &0.912$\pm0.033$
\\
\hline
$^4$He$\,${\protect\cite{chen1}}
        & 0.9   & $85^{\circ}$ & 0.316-0.696  & 1.617-0.256&
          0.959-0.335; & 0.78  & 5     & 1.014$\pm 0.030;$ &1.009$\pm0.029$
\\
        & 1.1   & $85^{\circ}$ & 0.492-0.796  & 1.322-0.408&
          1.221-0.610; & 1.09  & 5     & 0.994$\pm 0.062;$ &0.979$\pm0.062$
\\
\hline
 $R_T^{\rm He}$ & &            & 0.390-0.600& 1.298-0.659
       & 0.950-0.741; & 0.90      & 5 & 1.079$\pm 0.006 ;$&1.066$\pm0.006$
\\
\hline
   $D\,${\protect\cite{chen1,chen2}}
        & 4.045 & $15^{\circ}$ & 0.405-1.185  & 1.320-0.355&
          1.003-0.789; & 0.972 & 10    & 1.004$\pm 0.023;$ &0.945$\pm0.024$
\\
        & 4.045 & $23^{\circ}$ & 0.615-1.575  & 1.911-0.537&
          2.206-1.589; & 1.940 &  9    & 0.994$\pm 0.019;$ &0.921$\pm0.021$
\\
        & 4.045 & $30^{\circ}$ & 0.930-2.505  & 1.933-0.355&
          3.376-1.669; & 2.774 &  9    & 0.991$\pm 0.017;$ &0.915$\pm0.019$
\\
        & 4.045 & $37^{\circ}$ & 1.245-1.935  & 1.951-0.946&
          4.561-3.437; & 3.535 &  9    & 0.959$\pm 0.019;$ &0.874$\pm0.020$
\\
        & 4.045 & $45^{\circ}$ & 1.695-2.505  & 1.750-0.776&
          5.568-3.649; & 4.251 & 12    & 0.936$\pm 0.030;$ &0.845$\pm0.035$
\\
        & 4.045 & $55^{\circ}$ & 2.160-3.045  & 1.603-0.603&
          6.502-3.449; &4.900  &  7    & 0.879$\pm 0.018;$ &0.789$\pm0.018$
\\
\hline
 $D\,${\protect\cite{lung}}

        & 5.507 &  $15.15^{\circ}$ & 0.698-1.309 & 1.404-0.654&
          1.840-1.606; & 1.750 & 11    & 1.008$\pm 0.027;$ &0.933$\pm0.026$
\\
        & 5.507 &  $18.98^{\circ}$ & 1.011-1.665 & 1.418-0.736&
          2.692-2.300; & 2.500 & 17    & 0.998$\pm 0.039;$ &0.919$\pm0.040$
\\
        & 5.507 &  $22.81^{\circ}$ & 1.571-2.053 & 1.149-0.771&
          3.388-2.973; & 3.250 & 17    & 0.987$\pm 0.049;$ &0.900$\pm0.051$
\\
        & 5.507 &  $26.82^{\circ}$ & 1.983-2.424 & 1.121-0.803&
          4.176-3.653; & 4.000 &  6    & 0.951$\pm 0.058;$ &0.851$\pm0.059$
\\\hline
  $R_T^D     $ &   &               & 0.773-0.971 & 1.066-0.950&
          1.769-1.733; & 1.750 & 10    & 0.957$\pm 0.023;$ &0.940$\pm0.024$
\\
               &   &               & 1.281-1.394 & 1.051-0.940&
          2.529-2.461; & 2.500 & 5     & 0.973$\pm 0.022;$ &0.957$\pm0.022$
\\
               &   &               & 1.692-1.838 & 1.034-0.942&
          3.286-3.188; & 3.250 & 7     & 0.978$\pm 0.018;$ &0.959$\pm0.018$
\\
               &   &               & 2.095-2.210 & 1.027-0.941&
          4.059-3.905; & 4.000 & 4     & 0.901$\pm 0.025;$ &0.866$\pm0.025$
\\
\hline
 $D\,${\protect\cite{arrd}}
        & 2.015 &  $38.84^{\circ}$ & 0.582-0.722 & 1.168-0.850&
          1.277-1.152; & 1.22 &  7    & 0.924$\pm 0.033;$ &0.868$\pm0.036$ \\
        & 3.188 &  $47.68^{\circ}$ & 1.635-1.786 & 1.054-0.871&
          3.235-2.921; & 3.15 &  4    & 0.956$\pm 0.089;$ &0.897$\pm0.096$ \\
        & 4.212 &  $53.39^{\circ}$ & 2.659-2.810 & 1.058-0.904&
          5.280-4.767; & 5.10 &  6    & 0.855$\pm 0.064;$ &0.774$\pm0.070$ \\
        & 5.120 &  $56.64^{\circ}$ & 3.567-3.719 & 1.069-0.928&
          7.156-6.461; & 6.83 &  4    & 0.842$\pm 0.051;$ &0.786$\pm0.049$ \\
\hline
$D\,${\protect\cite{rock}}
        & 9.744  &  $10^{\circ}$      & 1.14-1.60  & 1.19-0.803&
          2.547-2.4112; & 2.50 &  6   & 0.977$\pm 0.056;$ &0.878$\pm0.061$ \\
        & 12.565 &  $10^{\circ}$      & 1.821-2.537 & 1.200-0.804&
          4.101-3.828;  & 4.00 &  7   & 0.913$\pm 0.050;$ &0.784$\pm0.053$ \\
        & 15.730 &  $10^{\circ}$      & 2.758-3.768 & 1.197-0.808&
          6.199-5.716;  & 6.00 &  10  & 0.854$\pm 0.029;$ &0.714$\pm0.031$ \\
        & 17.301 &  $10^{\circ}$      & 3.274-4.331 & 1.200-0.838&
          7.373-6.817; & 7.12 &  8   & 0.879$\pm 0.124;$ &0.812$\pm0.117$ \\
        & 18.476 &  $10^{\circ}$      & 3.087-4.748 & 1.191-0.864&
          8.289-7.705;  & 8.03 &  9   & 0.958$\pm 0.038;$ &0.912$\pm0.038$ \\
        & 20.999 &  $10^{\circ}$      & 4.658-5.885 & 1.192-0.873&
         10.424-9.642;  &10.06 &  5   & 0.769$\pm 0.118;$ &0.708$\pm0.131$ \\
\hline
\end{tabular}
\label{TableI}
}
\end{table}

\end{document}